\documentclass[12pt]{article}
\usepackage[english]{babel}
\usepackage{amsmath}
\usepackage{hyperref}
\begin{document}

\title{\bf{}Lagrangian formulation of massive fermionic totally antisymmetric
tensor field theory in $AdS_d$ space}

\author{\sc I.L. Buchbinder${}^{a}$\thanks{joseph@tspu.edu.ru},
V.A. Krykhtin${}^{a,b}$\thanks{krykhtin@tspu.edu.ru},
L.L. Ryskina$^a$\thanks{ryskina@tspu.edu.ru}
\\[0.5cm]
\it ${}^a$Department of Theoretical Physics,\\
\it Tomsk State Pedagogical University,\\
\it Tomsk 634061, Russia\\[0.3cm]
\it ${}^b$Laboratory of Mathematical Physics,\\
\it Tomsk Polytechnic University,\\
\it Tomsk 634050, Russia}

\date{}

\maketitle
\thispagestyle{empty}

\begin{abstract}
We apply the BRST approach, developed for higher spin field
theories, to Lagrangian construction for totally antisymmetric
massive fermionic fields in $AdS_{d}$ space. As well as generic
higher spin massive theories, the obtained Lagrangian theory is a
reducible gauge model containing, besides the basic field, a number
of auxiliary (St\"uckelberg) fields and the order of reducibility
grows with the value of the rank of the antisymmetric field.
However, unlike the generic higher spin theory, for the special case
under consideration we show that one can get rid of all the
auxiliary fields and the final Lagrangian for fermionic
antisymmetric field is formulated only in terms of basic field.
\end{abstract}

\section{Introduction}


One of the interesting aspects of higher spin field theory (see e.g.
the recent reviews \cite{reviews}) in various dimensions is a
possibility to construct the models using the fields with mixed
symmetry of tensor indices (see the recent papers
\cite{mixed1,mixed2,mixed3,mixed4}
and references therein). Since the
totally antisymmetric fields are particular cases of generic mixed
symmetry fields, the methods developed in higher spin field theory
can be applied for Lagrangian formulation of totally antisymmetric
fields.

In our recent paper \cite{08103467} we constructed the Lagrangians
for totally antisymmetric massive and massless bosonic fields in curved
space-time using BRST approach which was earlier applied for
description of totally symmetric higher spin field models
\cite{massless-bos,B-BRST-Ads,boz-ferm,0603212,0703049,0608005,massive-bos,0410215,INT}
and mixed symmetry higher spin fields
\cite{0101201}. In this paper we find the
Lagrangians for massive fermionic totally antisymmetric tensor
fields in AdS space using the BRST approach.

In principle, the Lagrangian for totally antisymmetric fermionic
field can be derived using the generic method developed in
\cite{0101201} for mixed symmetry fields.
This method uses the bosonic creation and annihilation operators which
automatically take into account symmetry of the group of the
indices. All the other symmetry conditions on the indices
are formulated in the form of constraints.
Such a procedure possesses a definite advanteges, however demands including a number of auxiliary
fields and gauge symetries in the Lagrangian.
In particular, for
totally antisymmetric fields  it means that
the antisymmetry index condition is absent from the very beginning and should be a
consequence of the equations of motion and gauge fixing.
Therefore a straightforward application of the generic method to totally
antisymmetric tensor fields being absolutely correct, yields complicated enough Lagrangian formulation.
In the paper under consideration we use fermionic creation and
annihilation operators thus taking into account the antisymmetry
index condition from the very beginning.
As a result the Lagrangian will
contain less number of the auxiliary fields and will be more simpler
in comparison with a result of straightforward application of the generic method
\cite{0101201}.\footnote {Of course, finally both
such ways lead to the same Lagrangian. Here we point out a
possibility to obtain the final result more simple way.}

The paper is organized as follows. In the next section we show that
the equations of motion for antisymmetric fermionic field are
noncontradictory for arbitrary dimensions only in spaces of constant
curvature if one supposes the absence of the terms with the inverse
powers of the mass. The rest part of the paper deals with the
fermionic fields on AdS background. In section~\ref{secAlgebra} we
rewrite the equations of motion for antisymmetric fermionic field in
the operator form and find the closed algebra generated by such
operators. Then according to the generic procedure of Lagrangian
construction \cite{0608005} we derive in section~\ref{secAddParts}
the additional parts and in section~\ref{secBRST} we first find the
extended expressions for the operators and then on the base of their
algebra we construct BRST operator. Finally in
section~\ref{secLagrangian} we determine the Lagrangian for the
antisymmetric fermionic field. In the Appendices we give some
details of calculations missed in the main part of the paper. In
appendix~\ref{AdParts} we describe the calculations of the
additional parts. In appendix~\ref{reduction} we show that the
obtained Lagrangian indeed reproduces the true equations determining
the irreducible representation of the $AdS_d$ group on massive
fermionic antisymmetric fields and in appendix~\ref{simple} we
simplify the Lagrangian by removing all the auxiliary fields and get
the final Lagrangian in terms of physical field only. It should be
noted that such Lagrangian has not been previously presented in
the literature.

\section{Consistency of fermionic field dynamics in curved space}

In this section we show that
unlike the bosonic case \cite{08103467} there are no consistent
equations of motion for fermionic totally antisymmetric fields
minimally coupled to arbitrary curved space-time.

It is well known that a rank-$n$ totally antisymmetrical
tensor-spinor field $\psi_{\mu_1\cdots\mu_n}$ (the Dirac index is
suppressed) will describe the irreducible massive representation of
the Poincare group if the following conditions are satisfied
\begin{eqnarray}
&&
(i\gamma^\nu\partial_\nu-m)\psi_{\mu_1\cdots\mu_n}=0,
\qquad
\gamma^\mu\psi_{\mu\mu_2\cdots\mu_n}=0,
\qquad
\partial^\mu\psi_{\mu\mu_2\cdots\mu_n}=0,
\label{irrep-f}
\end{eqnarray}
with $\{\gamma^\mu,\gamma^\nu\}=-2g^{\mu\nu}$.
When we put these equations on an arbitrary curved spacetime
we see that if we do not include the terms with the inverse powers of the
mass then
there is no freedom to add any terms with the curvature and it
unambiguously follows that in curved space equations
(\ref{irrep-f}) take the form
\begin{eqnarray}
(i\gamma^\nu\nabla_\nu-m)\psi_{\mu_1\cdots\mu_n}=0,
\qquad
\gamma^\mu\psi_{\mu\mu_2\cdots\mu_n}=0,
\qquad
\nabla^\mu\psi_{\mu\mu_2\cdots\mu_n}=0.
\label{irrep-fc}
\end{eqnarray}

Let us show that the mass-shell and divergence-free equations are
inconsistent in arbitrary curved space. For this purpose we take
the divergence of the mass-shell equation and suppose that equations
(\ref{irrep-fc}) are satisfied
\begin{eqnarray}
0&=&\nabla^{\mu_1}(i\gamma^\nu\nabla_\nu-m)\psi_{\mu_1\cdots\mu_n}
=i\gamma^\nu[\nabla^{\mu_1},\nabla_\nu]\psi_{\mu_1\cdots\mu_n}
\nonumber
\\
&=&
i\gamma^\nu\Bigl\{
R^\alpha_\nu\psi_{\alpha\mu_2\cdots\mu_n}
+R_{\mu_2}{}^{\alpha\mu_1}{}_\nu \psi_{\mu_1\alpha\mu_3\cdots\mu_n}
+\ldots
+R_{\mu_n}{}^{\alpha\mu_1}{}_\nu \psi_{\mu_1\cdots\mu_{n-1}\alpha}
-\frac{1}{4}R^{\alpha\beta\mu_1}{}_{\nu}\gamma_{\alpha\beta}\,\psi_{\mu_1\cdots\mu_n}
\Bigr\}
.
\label{compat}
\end{eqnarray}
One can see that the last expression in (\ref{compat}) assumes if
the space is arbitrary curved then $\psi_{\mu_1\cdots\mu_n}=0$. Let
us try to find from (\ref{compat}) what spaces do not give any
condition on $\psi_{\mu_1\cdots\mu_n}$. Let us decompose the Riemann
tensor
\begin{eqnarray}
R_{\mu\nu\alpha\beta}&=&
C_{\mu\nu\alpha\beta}
+\frac{1}{d-2}\left(
R_{\mu\alpha}g_{\nu\beta}+R_{\nu\beta}g_{\mu\alpha}
-R_{\mu\beta}g_{\nu\alpha}-R_{\nu\alpha}g_{\mu\beta}
\right)
\nonumber
\\
&&{}
+\frac{1}{(d-1)(d-2)}R(g_{\mu\beta}g_{\nu\alpha}-g_{\mu\alpha}g_{\nu\beta}),
\end{eqnarray}
where $C_{\mu\nu\alpha\beta}$ is the Weyl tensor,
and substitute this decomposition into (\ref{compat})
\begin{eqnarray}
i\gamma^\nu\Bigl\{
C_{\mu_2}{}^{\alpha\mu_1}{}_\nu \psi_{\mu_1\alpha\mu_3\cdots\mu_n}
+\ldots
+C_{\mu_n}{}^{\alpha\mu_1}{}_\nu \psi_{\mu_1\cdots\mu_{n-1}\alpha}
-\frac{1}{4}C^{\alpha\beta\mu_1}{}_{\nu}\gamma_{\alpha\beta}\,\psi_{\mu_1\cdots\mu_n}
\Bigr\}
\nonumber
\\
{}
+
\frac{i}{2}\;\frac{d-2n}{d-2}\;
R^{\alpha\mu_1}\gamma_\alpha\psi_{\mu_1\ldots\mu_n}
&=&0,
\label{compat-2}
\end{eqnarray}
where we have used $\gamma^{\mu_1}\psi_{\mu_1\ldots\mu_n}=0$. From
(\ref{compat-2}) we see that for compatibility of (\ref{irrep-fc})
one should suppose that $C_{\mu\nu\alpha\beta}=0$. After this
condition (\ref{compat-2}) is reduced to
\begin{eqnarray}
\frac{i}{2}\;\frac{d-2n}{d-2}\;
R^{\alpha\mu_1}\gamma_\alpha\psi_{\mu_1\ldots\mu_n}
&=&0.
\label{compat-3}
\end{eqnarray}
First we see that in even dimensional spaces when $d=2n$, with $n$
being the tensor rank of the field there is no restriction on the
Ricci tensor. This special case will be studied elsewhere. Now we
consider arbitrary values of $d$ and $n$. In this general case one
must suppose that the traceless part of the Ricci tensor is zero
$\tilde{R}_{\mu\nu}=R_{\mu\nu}-\frac{1}{d}g_{\mu\nu}R=0$, what means
that $R_{\mu\nu}=\frac{1}{d}g_{\mu\nu}R$. Next from the corollary of
the Bianci identity $2\nabla^\nu{}R_{\mu\nu}=\nabla_\mu{}R$ one
finds that $R=const$. That is equations (\ref{irrep-fc}) are
compatible with each other only on the spaces with constant
curvature. Therefore the rest part of the paper will be devoted to
Lagrangian construction for the fields in AdS space. Note that in
spaces of constant curvature there is a possibility to modify
equations of motion (\ref{irrep-fc}), since a parameter (the radius
of the curvature) with dimension of length appears.

\section{Algebra of constraints for fermionic fields in $AdS_d$}\label{secAlgebra}

As is known an antisymmetric tensor rank-$n$ fermionic
field will realize
irreducible massive
representation of the AdS group \cite{9802097} if the following conditions are
satisfied
\begin{eqnarray}
[i\gamma^\mu\nabla_\mu-m+r^\frac{1}{2}(n-{\textstyle\frac{d}{2}})]
\psi_{\mu_1\ldots\mu_n}=0,
\qquad
\gamma^{\mu_1}\psi_{\mu_1\ldots\mu_n}=0,
\qquad
\nabla^{\mu_1}\psi_{\mu_1\ldots\mu_n}=0.
\label{TheEM}
\end{eqnarray}
Here $r$ is defined from
$R_{\mu\nu\alpha\beta}=r(g_{\mu\beta}g_{\nu\alpha}-g_{\mu\alpha}g_{\nu\beta})$.
Analogously to the bosonic case \cite{08103467} in order
to avoid manipulations with a number of indices we
introduce auxiliary Fock space generated by fermionic creation and annihilation
operators $a_a^+$, $a_a$
satisfying the anticommutation relations
\begin{eqnarray}
\{a^+_a,a_b\}=\eta_{ab},
\qquad
\eta_{ab}=diag(-,+,+,\cdots,+)
.
\end{eqnarray}
As usual the tangent space indices and the curved indices are
converted one into another with the help of vielbein $e^a_\mu$
which is assumed to satisfy the relation $\nabla_\nu{}e^a_\mu=0$.
Then in addition to the conventional gamma-matrices
\begin{eqnarray}
\{\gamma_a, \gamma_b\}=-2\eta_{ab},
\end{eqnarray}
we introduce a set of $d+1$ Grassmann
odd objects \cite{0603212,0703049} which obey the following gamma-matrix-like conditions
\begin{eqnarray}
\{\tilde{\gamma}{}^a,\tilde{\gamma}{}^b\}=-2\eta^{ab},
&\qquad&
\{\tilde{\gamma}{}^a,\tilde{\gamma}\}=0,
\qquad
\tilde{\gamma}{}^2=-1
\label{tildedgamma}
\end{eqnarray}
and connected with
the ``true''
gamma-matrices by the relation
\begin{eqnarray}
\gamma^a&=&\tilde{\gamma}{}^a\tilde{\gamma}
           =-\tilde{\gamma}\tilde{\gamma}{}^a.
\label{truegamma}
\end{eqnarray}
After this we define derivative operator
\begin{eqnarray}
D_\mu=\partial_\mu+\omega_\mu{}^{ab}M_{ab},
\qquad
M_{ab}=
{\textstyle\frac{1}{2}}(a_a^+a_b-a_b^+a_a)
-{\textstyle\frac{1}{8}}(
\tilde{\gamma}_a\tilde{\gamma}_b-\tilde{\gamma}_b\tilde{\gamma}_a),
\end{eqnarray}
which acts on an arbitrary state vector in the Fock
space
\begin{eqnarray}
\label{PhysState}
|\psi\rangle&=&\sum_{n=0}
a^{+\mu_1}\cdots a^{+\mu_n}\psi_{\mu_1\cdots\mu_n}(x)|0\rangle
\end{eqnarray}
as the covariant derivative\footnote{We assume that
$\partial_\mu{}a_a^+=\partial_\mu{}a_a=\partial_\mu|0\rangle=0$.}
\begin{eqnarray}
D_\mu|\psi\rangle
&=&
\sum_{n=0}^{}
a^{\mu_1+}\ldots a^{\mu_n}
(\nabla_\mu\psi_{\mu_1\ldots\mu_n})|0\rangle.
\end{eqnarray}
As a result equations (\ref{TheEM}) can be realize in the operator form
\begin{eqnarray}
\label{TheEM-op}
\tilde{t}_0|\psi\rangle=0,
\qquad
t_1|\psi\rangle=0,
\qquad
l_1|\psi\rangle=0,
\end{eqnarray}
where
\begin{eqnarray}
\tilde{t}_0=i\tilde{\gamma}^\mu D_\mu
+\tilde{\gamma}(r^\frac{1}{2}g_0-m),
\qquad
g_0=a_\mu^+a^\mu-{\textstyle\frac{d}{2}},
\qquad
t_1=\tilde{\gamma}^\mu a_\mu,
\qquad
l_1=-ia^\mu D_\mu.
\end{eqnarray}

Lagrangian construction within the BRST approach \cite{0608005}
demands that we must have at hand a set of operators which is
invariant under Hermitian conjugation and which forms an algebra
\cite{0608005,massive-bos}. In order to determine the Hermitian
conjugation properties of the constraints we define the following
scalar product
\begin{equation}
\label{sproduct}
\langle\tilde{\Psi}|\Phi\rangle
=
\int d^dx \sqrt{-g} \sum_{n,\,k=0}
\langle0|
\Psi^+_{\nu_1\ldots\,\nu_k}(x) \tilde{\gamma}^0
a^{\nu_k}\ldots\,a^{\nu_{1}}
a^{+\mu_1}\ldots\,a^{+\mu_n}
\Phi_{\mu_1\ldots\,\mu_n}(x)
|0\rangle
.
\end{equation}
As a result we see that constraint $\tilde{t}_0$ is Hermitian and
the two other are non-Hermitian\footnote{We assume that
$(\tilde{\gamma}^\mu)^+=\tilde{\gamma}^0\tilde{\gamma}^\mu\tilde{\gamma}^0$,
$(\tilde{\gamma})^+=\tilde{\gamma}^0\tilde{\gamma}\tilde{\gamma}^0=-\tilde{\gamma}$.}
\begin{eqnarray}
t_1^+=a^+_\mu\tilde{\gamma}^\mu,
&\qquad&
l_1^+=-ia^{\mu+}D_\mu.
\end{eqnarray}
Now in order to have an algebra we add to the set of operators
all the operators generated by the (anti)commutators of
$\tilde{t}_0$, $t_1$, $l_1$, $t_1^+$, $l_1^+$.
Therefore we have to add the following three operators
\begin{eqnarray}
&&
\tilde{l}_0=D^2-m^2+r\bigl(-g_0^2+g_0+t_1^+t_1+{\textstyle\frac{d(d+1)}{4}}\bigr),
\\
&&
g_0=a_\mu^+a^\mu-{\textstyle\frac{d}{2}},
\qquad
g_m=m,
\end{eqnarray}
where
$D^2=g^{\mu\nu}(D_{\mu}D_\nu-\Gamma_{\mu\nu}^{\sigma}D_\sigma)$.
As a result set of operators
$\tilde{t}_0$, $\tilde{l_0}$, $t_1$, $l_1$, $t_1^+$, $l_1^+$,
$g_0$, $g_m$
is invariant under Hermitian conjugation and form an algebra.

The method of Lagrangian construction within the BRST approach
\cite{0608005} requires enlarging of the initial operators
\begin{math}
\tilde{o}_i=(\tilde{t}_0, \tilde{l_0}, t_1, l_1, t_1^+, l_1^+,
g_0, g_m)
\end{math}
so that the enlarged Hermitian operators contain arbitrary
parameters and the set of enlarged operators form an algebra. A
procedure of constructing of these enlarged operators
$\tilde{O}_i=\tilde{o}_i+\tilde{o}_i'$ for the operators
$\tilde{o}_i$ is considerably simplified if the initial operators
$\tilde{o}_i$ (super)commute with their additional parts\footnote{We
suppose that the additional parts are constructed from new
(additional) creation and annihilation operators and from the
constants of the theory. See e.g. \cite{0608005}.} $\tilde{o}_i'$:
$[\tilde{o}_i,\tilde{o}_j'\}=0$. In this case we can apply the
method elaborated in \cite{0608005}. If we try to construct the
enlarged operators $\tilde{O}_i=\tilde{o}_i+\tilde{o}_i'$  on the
base of initial operators
\begin{math}
\tilde{o}_i=(\tilde{t}_0, \tilde{l_0}, t_1, l_1, t_1^+, l_1^+,
g_0, g_m)
\end{math}
we find that the additional parts $\tilde{o}_i$ can't (super)commute
with the initial operators $\tilde{o}_i$. This happens because the
additional parts must contain $\tilde{\gamma}$ which is also present
in $\tilde{t}_0$. Therefore in order that initial operators
(super)commute with additional parts we make a non-degenerate linear
transformation ${o}_i =U^i_j\tilde{o}_j$ and remove $\tilde{\gamma}$
from $\tilde{t}_0$. Thus we modify $\tilde{t}_0$ and $\tilde{l}_0$
\begin{align}
&
t_0=\tilde{t}_0+\tilde{\gamma}(g_m-r^\frac{1}{2}g_0),
&&
l_0=\tilde{l}_0+g_m^2+rg_0^2,
\\
&
t_0=i\tilde{\gamma}^\mu D_\mu,
&&
l_0=D^2+r(g_0+t_1^+t_1+{\textstyle\frac{d(d+1)}{4}})
\end{align}
with the other operators being unchanged.
Due to this transformation of the initial operators we can apply
the method of constructing of additional parts elaborated in
\cite{0608005}. Algebra of new initial operators is given in
Table~\ref{table}
\begin{table}
\begin{center}
\begin{tabular}{|l||c|c|c|c|c|c|c|c||}
  \hline
 $\left[\; \downarrow, \rightarrow \right\}$  & $t_0$ & $t_1$  &
            $t_1^+$  &  $l_0$  &   $l_1$  &  $l_1^+$ & $g_0$ & $g_m$\\
  \hline
  \hline
  $t_0$
  &$2l_0$&$2l_1$&$-2l_1^+$&0&(\ref{at0l1})&(\ref{at0l1+})&0&0\\
  \hline
  $t_1$  &$-2l_1$&0&$2g_0$&(\ref{at1l0})&0&$-t_0$&$t_1$&0  \\
  \hline
  $t_1^+$  &$2l_1^+$&$-2g_0$&0&(\ref{at1+l0})&$t_0$&0&$-t_1^+$&0\\
  \hline
  $l_0$  &0&$-$(\ref{at1l0})&$-$(\ref{at1+l0})& 0
  &(\ref{al0l1})&(\ref{al0l1+})&0&0\\
  \hline
  $l_1$
  &(\ref{at0l1})&0&$-t_0$&$-$(\ref{al0l1})&$\frac{1}{2}rt_1^2$&(\ref{al1l1+})
  &$l_1$&0\\
  \hline
  $l_1^+$
  &(\ref{at0l1+})&$t_0$&0&$-$(\ref{al0l1+})&(\ref{al1l1+})&$\frac{1}{2}rt_1^{+2}$&$-l_1^+$&0\\
  \hline
  $g_0$ &0&$-t_1$&$t_1^+$&0&$-l_1$&$l_1^+$&0&0\\
  \hline
$g_m$&0&0&0&0&0&0&0&0\\
  \hline
  \hline
\end{tabular}
\end{center}
\caption{Algebra of the initial operators}
\label{table}
\end{table}
with
\begin{eqnarray}
\{t_0,l_1\}
&=&
-r(g_0+{\textstyle\frac{1}{2}})t_1,
\label{at0l1}
\\
\{t_0,l_1^+\}&=&-r\,t_1^+(g_0+{\textstyle\frac{1}{2}}),
\label{at0l1+}
\\
{}
[t_1,l_0]&=&
r(2g_0+1)t_1,
\label{at1l0}
\\
{}
[t_1^+,l_0]&=&
-r\,t_1^+(2g_0+1),
\label{at1+l0}
\\
{}
[l_0,l_1]&=&-r(2g_0+1)l_1,
\label{al0l1}
\\
{}
[l_0,l_1^+]&=&r\,l_1^+(2g_0+1),
\label{al0l1+}
\\
\{l_1,l_1^+\}
&=&
-l_0
+r(g_0^2+{\textstyle\frac{1}{2}}g_0+{\textstyle\frac{1}{2}}t_1^+t_1).
\label{al1l1+}
\end{eqnarray}

Next step in the procedure of Lagrangian construction is finding
the additional parts for the initial operators given in
Table~\ref{table}.

\section{The additional parts}\label{secAddParts}

In this section we are going to find explicit expressions for
the additional parts in terms of new (additional) creation and
annihilation operators and from the constants of the theory.
The requirements the additional parts $o_i'$ must satisfy is as
follows: 1)~The enlarged operators $O_i=o_i+o_i'$ are in
involution relation $[O_i,O_j]\sim{}O_k$; 2)~each Hermitian
operator must contain an arbitrary parameter linearly which
values shall be defined later from the condition of reproducing
the equation of motion (\ref{TheEM}).

To find explicit expression for the additional parts we must
first determine their algebra. The procedure of finding the
algebra of the additional parts for nonlinear algebras
was elaborated in \cite{0608005}.
To be complete we explain this procedure using anticommutator
$\{L_1,L_1^+\}$ as an example. Supposing that the initial
operators $o_i$ (super)commute with the additional parts $o_i'$
one finds
\begin{eqnarray}
\{L_1, L_1^+\}&=&\{l_1, l_1^+\}+\{l_1', l_1^{\prime+}\}
=
-l_0
+r(g_0^2+{\textstyle\frac{1}{2}}g_0+{\textstyle\frac{1}{2}}t_1^+t_1)
+\{l_1', l_1^{\prime+}\}
.
\end{eqnarray}
Then we express all the initial operators through the enlarged and
the additional ones $o_i=O_i-o_i'$ and order the operators so
that the enlarged operators stand on the right side
\begin{eqnarray}
\{L_1.L_1^+\}&=&
-L_0
+rG_0^2-2rg_0'G_0+{\textstyle\frac{1}{2}}rG_0
+{\textstyle\frac{1}{2}}rT_1^+T_1
-{\textstyle\frac{1}{2}}rt_1^{\prime+}T_1
-{\textstyle\frac{1}{2}}rt_1'T_1^+
\nonumber
\\
&&{}
+l_0'
+r(g_0^{\prime2}+{\textstyle\frac{1}{2}}g_0'+{\textstyle\frac{1}{2}}t_1^{\prime+}t_1')
+\{l_1', l_1^{\prime+}\}
.
\end{eqnarray}
In order to satisfy the first requirement $[O_i,O_j]\sim{}Q_k$ we put
\begin{eqnarray}
\{l_1',l_1^{\prime+}\}
&=&
-l_0'
-r(g_0^{\prime2}+{\textstyle\frac{1}{2}}g_0'+{\textstyle\frac{1}{2}}t_1^{\prime+}t_1')
\label{adl1l1+}
\end{eqnarray}
and as a consequence we get
\begin{eqnarray}
\{L_1,L_1^+\}
&=&
-L_0
+rG_0^2-2rg_0'G_0+{\textstyle\frac{1}{2}}rG_0
+{\textstyle\frac{1}{2}}rT_1^+T_1
-{\textstyle\frac{1}{2}}rt_1^{\prime+}T_1
-{\textstyle\frac{1}{2}}rt_1'T_1^+
.
\label{L1L1+}
\end{eqnarray}
Thus we have found anticommutator for the additional parts
$\{l_1',l_1^{\prime+}\}$ (\ref{adl1l1+}) and simultaneously anticommutator for the
enlarged
operators $\{L_1,L_1^+\}$ (\ref{L1L1+}).
Repeating the same procedure for the other (anti)commutators we
find the algebra of the additional parts and the algebra of the
extended operators.
The algebra of the additional parts
is
given\footnote{The algebra of the extended operators will be
discussed later. It is given in Table~\ref{Table} at
page~\pageref{Table}.}
in
Table~\ref{table'}
\begin{table}
\begin{center}
\begin{tabular}{|l||c|c|c|c|c|c|c|c||}
  \hline
 $\left[\; \downarrow, \rightarrow \right\}$  & $t_0'$ & $t_1'$  &
            $t_1^{\prime+}$&$l_0'$&$l_1'$&$l_1^{\prime+}$ &
            $g_0'$ & $g_m'$ \\
  \hline
  \hline
  $t_0'$
  &$2l_0'$&$2l_1'$&$-2l_1^{\prime+}$&0&(\ref{adt0l1})&(\ref{adt0l1+})&0&0\\
  \hline
  $t_1'$  &$-2l_1'$&0&$2g_0'$&(\ref{adt1l0})&0&$-t_0'$&$t_1'$ &0 \\
  \hline
  $t_1^{\prime+}$&$2l_1^{\prime+}$&$-2g_0'$&0&(\ref{adt1+l0})&$t_0'$&0&$-t_1^{\prime+}$&0\\
  \hline
  $l_0'$  &0&$-$(\ref{adt1l0})&$-$(\ref{adt1+l0})& 0
  &(\ref{adl0l1})&(\ref{adl0l1+})&0&0\\
  \hline
  $l_1'$
  &(\ref{adt0l1})&0&$-t_0'$&$-$(\ref{adl0l1})&$-\frac{1}{2}rt_1^{\prime2}$&(\ref{adl1l1+})
  &$l_1'$&0\\
  \hline
  $l_1^{\prime+}$
  &(\ref{adt0l1+})&$t_0'$&0&$-$(\ref{adl0l1+})&(\ref{adl1l1+})&$-\frac{1}{2}rt_1^{\prime+2}$&$-l_1^{\prime+}$&0\\
  \hline
  $g_0'$
  &0&$-t_1'$&$t_1^{\prime+}$&0&$-l_1'$&$l_1^{\prime+}$&0&0\\
  \hline
  $g_m'$&0&0&0&0&0&0&0&0\\
  \hline
  \hline
\end{tabular}
\end{center}
\caption{Algebra of the additional parts}
\label{table'}
\end{table}
with
\begin{eqnarray}
\{t_0',l_1'\}
&=&
r(g_0'+{\textstyle\frac{1}{2}})t_1',
\label{adt0l1}
\\
\{t_0',l_1^{\prime+}\}&=&r\,t_1^{\prime+}(g_0'+{\textstyle\frac{1}{2}}),
\label{adt0l1+}
\\
{}
[t_1',l_0']&=&-r(2g_0'+1)t_1',
\label{adt1l0}
\\
{}
[t_1^{\prime+},l_0']&=&
r\,t_1^{\prime+}(2g_0'+1),
\label{adt1+l0}
\\
{}
[l_0',l_1']&=&r(2g_0'+1)l_1',
\label{adl0l1}
\\
{}
[l_0',l_1^{\prime+}]&=&-r\,l_1^{\prime+}(2g_0'+1)
.
\label{adl0l1+}
\end{eqnarray}

Using this algebra one can find explicit expressions for the
additional parts in terms of new (additional) creation and
annihilation operators.
The method which allows us to do this is described in
appendix~\ref{AdParts}. The result takes the form
\begin{eqnarray}
\label{ad-1}
&&
t_1^{\prime+}=b^+,
\qquad
l_1^{\prime+}=m_1f^+-\frac{r}{4m_1}\,b^{+2}f,
\qquad
g_0'=b^+b+f^+f+h,
\qquad
g_m=h_m,
\\
&&
t_0'=-\tilde{\gamma}m_0
-2m_1f^+b
+\frac{r}{2m_1}\Bigl(b^+b+2h\Bigr)b^+f,
\\
&&
l_0'=-m_0^2-r(b^+b+2h)b^+b-2r(b^+b+h+{\textstyle\frac{1}{2}})f^+f,
\\
&&
t_1'=\tilde{\gamma}\frac{m_0}{m_1}f+(2f^+f+b^+b+2h)b,
\\
&&
l_1'=\tilde{\gamma}m_0b+m_1f^+b^2+\frac{m_0^2}{m_1}f
-\frac{r}{m_1}(h+{\textstyle\frac{1}{2}})(b^+b+h)f
-\frac{r}{4m_1}b^{+2}b^2f
,
\label{ad-5}
\end{eqnarray}
where we have introduced one pair of fermionic $f^+$, $f$ and
one pair of bosonic $b^+$, $b$ creation and annihilation
operators with the standard (anti)commutation relations
\begin{eqnarray}
\{f^+,f\}=1,&\qquad& [b^+,b]=1.
\end{eqnarray}
According to the second requirement the found additional parts
for Hermitian initial operators contain arbitrary parameters linearly:
operators $t_0'$, $g_0'$, $g_m'$ contain parameters $m_0$, $h$,
$h_m$ respectively.
Operator $l_0'$ cannot contain independent arbitrary parameter
since $l_0'=(t_0')^2$.
Parameters $m_0$ and $h_m$ have dimension of mass, and parameter
$h$ is dimensionless.
The values of these parameters will be defined later from the
condition of reproducing equations of motion (\ref{TheEM}).
Also expressions for additional parts (\ref{ad-1})--(\ref{ad-5})
contain arbitrary (nonzero) parameter $m_1$ with dimension of mass. Its value remains
arbitrary and it can be expressed from the other parameters of
the theory $m_1=f(m,r)\neq0$. The arbitrariness of this
parameter does not influence on the reproducing of the equations
of motion for the physical field (\ref{TheEM}).

Note that the additional parts do not obey the usual properties
\begin{align}
&
(t_0')^+\neq t_0'
&&
(l_0')^+\neq l_0',
&&
(t_1')^+\neq t_1^{\prime+},
&&
(l_1')^+\neq l_1^{\prime+}
\end{align}
if one use the standard rules of Hermitian conjugation for the new
creation and annihilation operators
\begin{equation}
(b)^+=b^+,
\qquad
(f)^+=f^+.
\end{equation}
To restore the proper Hermitian conjugation properties for the
additional parts we change the scalar product in the Fock space generated by the new creation and annihilation
operators  as follows:
\begin{eqnarray}
\langle\tilde{\Psi}_1|\Psi_2\rangle_{\mathrm{new}} =
\langle\tilde{\Psi}_1|K|\Psi_2\rangle\,, \label{newsprod}
\end{eqnarray}
for any vectors $|\Psi_1\rangle, |\Psi_2\rangle$ with some yet
unknown operator $K$. This operator is determined by the
condition that all the operators of the algebra  must have the
proper Hermitian properties with respect to the new scalar product:
\begin{align}
\label{H0}
&
\langle\tilde{\Psi}_1|Kt_0'|\Psi_2\rangle =
\langle\tilde{\Psi}_2|Kt_0'|\Psi_1\rangle^* ,
&&
\langle\tilde{\Psi}_1|Kl_1'|\Psi_2\rangle =
\langle\tilde{\Psi}_2|Kl_1^{\prime+}|\Psi_1\rangle^* ,
\\
&
\langle\tilde{\Psi}_1|Kl_0'|\Psi_2\rangle =
\langle\tilde{\Psi}_2|Kl_0'|\Psi_1\rangle^* ,
&&
\langle\tilde{\Psi}_1|Kt_1'|\Psi_2\rangle =
\langle\tilde{\Psi}_2|Kt_1^{\prime+}|\Psi_1\rangle^* ,
\\
&
\langle\tilde{\Psi}_1|Kg_0'|\Psi_2\rangle =
\langle\tilde{\Psi}_2|Kg_0'|\Psi_1\rangle^*
&&
\langle\tilde{\Psi}_1|Kg_m'|\Psi_2\rangle =
\langle\tilde{\Psi}_2|Kg_m'|\Psi_1\rangle^*
\label{H2}
.
\end{align}
Since the problem with the proper Hermitian conjugation of the operators are
in $(b^+, f^+)$-sector of the Fock space then the modification of the
scalar product concerns only this sector.
Therefore operator $K$ acts as a unit operator in the entire Fock space, but
for the $(b^+, f^+)$-sector where the operator has the form
\begin{eqnarray}
K&=&\sum_{k=0}^{\infty}\frac{C_h(k)}{k!}\Biggl[\;
|0,k\rangle\frac{1}{2h+k}\langle0,k|
\;+\;|1,k\rangle\frac{2m_0^2-rh(2h{+}k{+}1)}{4h\,m_1^2}\langle1,k|
\nonumber
\\
&&\hspace*{7em}{}
+|1,k\rangle\frac{\tilde{\gamma}m_0}{2h\,m_1}\langle0,k{+}1|
\;+\;|0,k{+}1\rangle\frac{\tilde{\gamma}m_0}{2h\,m_1}\langle1,k|
\;\;\Biggr]
,
\label{K}
\end{eqnarray}
where
\begin{eqnarray}
&&
C_h(k)=2h\,(2h+1)\cdot\ldots\cdot(2h+k-2)(2h+k-1)(2h+k),
\\
&&
|0,k\rangle=(b^+)^k|0\rangle,
\qquad
|1,k\rangle=f^+(b^+)^k|0\rangle.
\end{eqnarray}

Thus in this section we have constructed  the additional parts
(\ref{ad-1})--(\ref{ad-5}) for the operators which obey all
the requirements. In the next section we determine the algebra
of the enlarged operators and construct BRST operator
corresponding to this algebra.

\section{The deformed algebra and BRST operator}\label{secBRST}

The algebra of the enlarged operators can be determined by the
method described in the previous section where we obtained
anticommutator $\{L_1,L_1^+\}$ in the explicit form (\ref{L1L1+}).
Looking at this anticommutator we see that its r.h.s. are
quadratic and therefore there is different possibilities to
order operators
\begin{eqnarray}
\{L_1,L_1^+\}
&=&
-L_0
+rG_0^2-2rg_0'G_0
-r({\textstyle\frac{1}{2}}-\xi)G_0
\nonumber
\\
&&{}
+{\textstyle\frac{1}{2}}r\xi T_1^+T_1
+{\textstyle\frac{1}{2}}r(1-\xi)T_1T_1^+
-{\textstyle\frac{1}{2}}rt_1^{\prime+}T_1
-{\textstyle\frac{1}{2}}rt_1'T_1^+,
\end{eqnarray}
where we have introduced parameter $\xi$ responsible for the
operator ordering.
The same is valid for the other (anti)commutators.
Each ordering leads to different forms of BRST operator.
We will not investigate all the possibilities of ordering here
and choose only one of them which corresponds to the
supersymmetric ordering of the enlarged operators in the rhs.
The
algebra\footnote{We put arbitrary constant
$h_m$ to $-m$ and get that enlarged operator $G_m=g_m+g_m'=0$. In
what follows we forget about $G_m$.}
of
the enlarged operators corresponding to the
supersymmetric ordering is presented in Table~\ref{Table}
\begin{table}
\begin{center}
\begin{tabular}{|l||c|c|c|c|c|c|c||}
  \hline
 $\left[\; \downarrow, \rightarrow \right\}$  & $T_0$ & $T_1$  &
            $T_1^+$  &  $L_0$  &   $L_1$  &  $L_1^+$ & $G_0$ \\
  \hline
  \hline
  $T_0$  &$2L_0$&$2L_1$&$-2L_1^+$&0&(\ref{et0l1})&(\ref{et0l1+})&0\\
  \hline
  $T_1$  &$-2L_1$&0&$2G_0$&(\ref{et1l0})&0&$-T_0$&$T_1$  \\
  \hline
  $T_1^+$  &$2L_1^+$&$-2G_0$&0&(\ref{et1+l0})&$T_0$&0&$-T_1^+$\\
  \hline
  $L_0$  &0&$-$(\ref{et1l0})&$-$(\ref{et1+l0})&0&(\ref{el0l1})&(\ref{el0l1+})&0\\
  \hline
  $L_1$   &(\ref{et0l1})&0&$-T_0$&$-$(\ref{el0l1})&(\ref{el1l1})&(\ref{el1l1+})&$L_1$\\
  \hline
  $L_1^+$   &(\ref{et0l1+})&$T_0$&0&$-$(\ref{el0l1+})&(\ref{el1l1+})&(\ref{el1+l1+})&$-L_1^+$\\
  \hline
  $G_0$ &0&$-T_1$&$T_1^+$&0&$-L_1$&$L_1^+$&0\\
  \hline
  \hline
\end{tabular}
\end{center}
\caption{Algebra of the enlarged operators}
\label{Table}
\end{table}
where
\begin{eqnarray}
\{T_0,L_1\}
&=&
-{\textstyle\frac{1}{2}}rG_0T_1
-{\textstyle\frac{1}{2}}rT_1G_0
+rg_0'T_1+rt_1'G_0,
\label{et0l1}
\\
\{T_0,L_1^+\}&=&
-{\textstyle\frac{1}{2}}r\,T_1^+G_0
-{\textstyle\frac{1}{2}}r\,G_0T_1^+
+rt_1^{\prime+}G_0+rg_0'T_1^+
,
\label{et0l1+}
\\
{}
[T_1,L_0]&=&rG_0T_1+rT_1G_0-2rg_0'T_1-2rt_1'G_0
,
\label{et1l0}
\\
{}
[T_1^+,L_0]&=&-rT_1^+G_0-rG_0T_1^++2rg_0'T_1^++2rt_1^{\prime+}G_0,
\label{et1+l0}
\\
{}
[L_0,L_1]&=&-rG_0L_1-rL_1G_0+2rg_0'L_1+2rl_1'G_0,
\label{el0l1}
\\
{}
[L_0,L_1^+]&=&rL_1^+G_0+rG_0L_1^+-2rl_1^{\prime+}G_0-2rg_0'L_1^+,
\label{el0l1+}
\\
\{L_1,L_1^+\}
&=&
-L_0
+rG_0^2-2rg_0'G_0
+{\textstyle\frac{1}{4}}rT_1^+T_1
+{\textstyle\frac{1}{4}}rT_1T_1^+
-{\textstyle\frac{1}{2}}rt_1^{\prime+}T_1
-{\textstyle\frac{1}{2}}rt_1'T_1^+,
\label{el1l1+}
\\
\{L_1,L_1\}&=&
{\textstyle\frac{1}{2}}rT_1^2-rt_1'T_1,
\label{el1l1}
\\
\{L_1^+,L_1^+\}&=&{\textstyle\frac{1}{2}}rT_1^{+2}-rt_1^{\prime+}T_1^+
.
\label{el1+l1+}
\end{eqnarray}

The construction of a nilpotent fermionic BRST operator for a
nonlinear  superalgebra
is based on the same principles as those developed in
\cite{B-BRST-Ads, 0608005} (for a general consideration of
operator BFV quantization, see the reviews \cite{bf}). The BRST operator
constructed on a basis of the algebra given by
Table~\ref{Table} is
\begin{eqnarray}
Q'&=&
q_0T_0+\eta_1^+T_1+\eta_1T_1^++\eta_0L_0+q_1^+L_1+q_1L_1^++\eta_GG_0
+(q_1^+q_1-q_0^2)\mathcal{P}_0
\nonumber
\\
&&{}
+2i\eta_1^+q_0p_1-2iq_0\eta_1p_1^+
-2\eta_1^+\eta_1\mathcal{P}_G
+i(\eta_1^+q_1-q_1^+\eta_1)p_0
\nonumber
\\
&&{}
+\eta_G(\eta_1^+\mathcal{P}_1-\eta_1\mathcal{P}_1^+
+iq_1^+p_1-iq_1p_1^+)
-rq_1^+q_1
(G_0-2g_0')\mathcal{P}_G
\nonumber
\\
&&{}
-r(\eta_1^+\eta_0-{\textstyle\frac{1}{2}}q_1^+q_0)\Bigl[
(G_0-2g_0')\mathcal{P}_1
+(T_1-2t_1')\mathcal{P}_G
\Bigr]
\nonumber
\\
&&{}
-r(\eta_0\eta_1-{\textstyle\frac{1}{2}}q_0q_1)\Bigl[
(G_0-2g_0')\mathcal{P}_1^+
+(T_1^+-2t_1^{\prime+})\mathcal{P}_G
\Bigr]
\nonumber
\\
&&{}
+rq_1^+\eta_0\Bigl[
(G_0-2g_0')ip_1
+(L_1-2l_1')\mathcal{P}_G
\Bigr]
\nonumber
\\
&&{}
-r\eta_0q_1\Bigl[
(G_0-2g_0')ip_1^+
+(L_1^+-2l_1^{\prime+})\mathcal{P}_G
\Bigr]
\nonumber
\\
&&{}
-\frac{r}{4}\Bigl[
q_1(T_1^+-2t_1^{\prime'})+q_1^+(T_1-2t_1')
\Bigr]
(q_1\mathcal{P}_1^++q_1^+\mathcal{P}_1)
\nonumber
\\
&&{}
+\frac{r^2}{4}\eta_0(q_1T_1^+-q_1^+T_1)
(q_1\mathcal{P}_1^++q_1^+\mathcal{P}_1)\mathcal{P}_G
.
\label{Q'}
\end{eqnarray}
Here, $q_0$, $q_1$, $q_1^+$ and  $\eta_0$, $\eta_1^+$, $\eta_1$, $\eta_G$ are, respectively, the bosonic and
fermionic ghost ``coordinates'' corresponding to their canonically
conjugate ghost ``momenta'' $p_0$, $p_1^+$, $p_1$, ${\cal{}P}_0$,
${\cal{}P}_1$, ${\cal{}P}_1^+$, ${\cal{}P}_G$. They obey the (anti)commutation relations
\begin{eqnarray}
\{\eta_0,\mathcal{P}_0\}=
\{\eta_G,\mathcal{P}_G\}=
\{\eta_1,\mathcal{P}_1^+\}=
\{\eta_1^+,\mathcal{P}_1\}=1,
&\quad&
[q_0,p_0]=[q_1,p_1^+]=[q_1^+,p_1]=i
\label{ghosts}
\end{eqnarray}
and possess the standard  ghost number distribution,
$gh(\mathcal{C}^i)=-gh(\mathcal{P}_i)=1$,
providing $gh(\tilde{Q}')=1$.
The  resulting BRST operator $Q'$ is Hermitian with respect to the
new scalar product (\ref{newsprod}).
Let us turn to Lagrangian construction on the base of BRST
operator $Q'$ (\ref{Q'}).

\section{Construction of Lagrangians}\label{secLagrangian}

In this section we construct Lagrangians of antisymmetric fermionic massive
fields in the AdS space.
This construction goes along the line of
\cite{0410215}.
First we extract the dependence of the BRST
operator $Q'$ (\ref{Q'}) on the ghosts $\eta_G$, $\mathcal{P}_G$
\begin{eqnarray}
Q'&=&Q+\eta_G(\sigma+h)+A\mathcal{P}_G
\\
Q&=&
q_0\Bigl[T_0
+2i(\eta_1^+p_1-\eta_1p_1^+)
+\frac{r}{2}(G_0-2g_0')(q_1\mathcal{P}_1^++q_1^+\mathcal{P}_1)
\Bigr]
+i(\eta_1^+q_1-q_1^+\eta_1)p_0
\nonumber
\\
&&{}
+\eta_0\Bigl[L_0
+r(G_0-2g_0')(
\eta_1^+\mathcal{P}_1-\eta_1\mathcal{P}_1^+
+iq_1^+p_1-iq_1p_1^+
)\Bigr]
+(q_1^+q_1-q_0^2)\mathcal{P}_0
\nonumber
\\
&&{}
+\eta_1^+T_1+\eta_1T_1^+
+q_1^+L_1+q_1L_1^+
\nonumber
\\
&&{}
-\frac{r}{4}\Bigl[
q_1(T_1^+-2t_1^{\prime'})+q_1^+(T_1-2t_1')
\Bigr]
(q_1\mathcal{P}_1^++q_1^+\mathcal{P}_1)
\\
\sigma+h&=&G_0
+\eta_1^+\mathcal{P}_1-\eta_1\mathcal{P}_1^+
+iq_1^+p_1-iq_1p_1^+
\end{eqnarray}
where explicit expression for the operator $A$ is not
essential.
Then we choose the following representation of the Hilbert space:
\begin{equation}
(p_0, q_1, p_1, \mathcal{P}_0, \mathcal{P}_G, \eta_1, \mathcal{P}_1 )
|0\rangle
=0
\label{ghostvac}
\end{equation}
and suppose
that the  vectors and gauge parameters do not depend on
$\eta_G$ \cite{0410215}
\begin{eqnarray}
\label{chi}
|\chi\rangle
&=&
\sum\limits_{k_i}
(q_0)^{k_1}(q_1^+)^{k_2}(p_1^+)^{k_3}(\eta_0)^{k_4}
(\eta_1^+)^{k_5}(\mathcal{P}_1^+)^{k_6}(b^+)^{k_{7}}(f^+)^{k_8}
a^{+{}\mu_1}\cdots a^{+{}\mu_{k_0}}\chi^{k_1 \cdots k_{8}}_{\mu_1\cdots \mu_{k_0}}(x)
|0\rangle.
\end{eqnarray}
The sum in (\ref{chi}) is taken  over $k_0, k_1, k_2$, $k_3$,
$k_7$ running from 0 to infinity and over $k_4, k_5,
k_6, k_8$ running from 0 to 1.
Then we derive from the
equation on the physical vector
$Q'|\chi\rangle=0$
and from the reducible gauge transformations
$\delta|\chi\rangle=Q'|\Lambda\rangle$ a sequence of relations:
\begin{align}
\label{Qchi}
&  Q|\chi\rangle=0,
&& (\sigma+h)|\chi\rangle=0,
&& gh(|\chi\rangle)=0,
\\
&  \delta|\chi\rangle=Q|\Lambda\rangle,
&& (\sigma+h)|\Lambda\rangle=0,
&& gh(|\Lambda\rangle)=-1,
\label{QLambda}
\\
&  \delta|\Lambda\rangle=Q|\Lambda^{(1)}\rangle,
&& (\sigma+h)|\Lambda^{(1)}\rangle=0,
&& gh(|\Lambda^{(1)}\rangle)=-2,
\\
&  \delta|\Lambda^{(i-1)}\rangle=Q|\Lambda^{(i)}\rangle,
&& (\sigma+h)|\Lambda^{(i)}\rangle=0,
&& gh(|\Lambda^{(i)}\rangle)= -(i+1).
\label{QLambdai}
\end{align}
The middle equation in (\ref{Qchi})
presents the equations for the possible values of $h$
\begin{eqnarray}
\label{h}
h&=&\frac{d}{2}-n,
\end{eqnarray}
with $n$ being related to the tensor rank of antisymmetric
tensor-spinor.
By fixing the tensor rank of the antisymmetric field
we also fix the parameter $h$ according to
(\ref{h}).
Having fixed a value of $h$ we should substitute it into each of
the expressions (\ref{Qchi})--(\ref{QLambdai}),
see \cite{0410215} for more details.

Next step is to extract the zero ghost mode from the
operator $Q$.
This operator has the structure
\begin{equation}
\label{0Q}
Q=
\eta_0\tilde{L}_0
+(q_1^+q_1-q_0^2){\cal{}P}_0
+q_0\tilde{T}_0
+i(\eta_1^+q_1-\eta_1q_1^+)p_0
+\Delta{}Q,
\end{equation}
where $\tilde{T}_0$, $\tilde{L}_0$, $\Delta{}Q$ is independent of
$\eta_0$, ${\cal{}P}_0$, $q_0$, $p_0$
\begin{eqnarray}
\tilde{T}_0&=&
T_0
+2i(\eta_1^+p_1-\eta_1p_1^+)
+\frac{r}{2}(G_0-2g_0')(q_1\mathcal{P}_1^++q_1^+\mathcal{P}_1)
\\
\tilde{L}_0&=&
L_0
+r(G_0-2g_0')(
\eta_1^+\mathcal{P}_1-\eta_1\mathcal{P}_1^+
+iq_1^+p_1-iq_1p_1^+)
\\
\Delta{}Q&=&
\eta_1^+T_1+\eta_1T_1^+
+q_1^+L_1+q_1L_1^+
-\frac{r}{4}\Bigl[
q_1(T_1^+-2t_1^{\prime+})+q_1^+(T_1-2t_1')
\Bigr]
(q_1\mathcal{P}_1^++q_1^+\mathcal{P}_1)
\end{eqnarray}
Also we decompose the state vector and the gauge parameters as
\begin{align}
\label{0chi}
|\chi\rangle
&=\sum_{k=0}^{\infty}q_0^k(
|\chi_0^k\rangle
+\eta_0|\chi_1^k\rangle),
&
&gh(|\chi^{k}_{m}\rangle)=-(m+k),
\\
\label{0L}
|\Lambda^{(i)}\rangle
&=\sum_{k=0}^{\infty}q_0^k(|\Lambda^{(i)}{}^k_0\rangle
+\eta_0|\Lambda^{(i)}{}^k_1\rangle),
&
&gh(|\Lambda^{(i)}{}^k_m\rangle)=-(i+k+m+1)
.
\end{align}
Then
following the procedure described in \cite{0410215}
we get rid of all the
fields except two $|\chi^0_0\rangle$, $|\chi^1_0\rangle$ and
the leftmost equation in (\ref{Qchi}) is reduced to
\begin{eqnarray}
&&
\Delta{}Q|\chi^{0}_{0}\rangle
+\frac{1}{2}\bigl\{\tilde{T}_0,q_1^+q_1\bigr\}|\chi^{1}_{0}\rangle
=0,
\label{EofM1all}
\\&&
\tilde{T}_0|\chi^{0}_{0}\rangle
+
\Delta{}Q|\chi^{1}_{0}\rangle
=0,
\label{EofM2all}
\end{eqnarray}
where
$\{A,B\}=AB+BA$.
State vector (\ref{chi}) and as a consequence
$|\chi_0^0\rangle$, $|\chi_0^1\rangle$ (\ref{0chi})
contain
physical\footnote{The physical fields in (\ref{chi}) are those
which correspond to $k_1=k_2=k_3=k_4=k_5=k_6=k_7=k_8=0$ and
arbitrary $k_0$ which is equal to $n$ being the tensor rank of
tensor-spinor. The other fields in decomposition (\ref{chi}) are
St\"uckelberg ($k_8=1$) or auxiliary ($k_8=0$).}
fields of all ranks.
Due to the fact that the  operators $\Delta{}Q$, $\tilde{T}_0$,
$q_1^+q_1$ commute with  $\sigma$ we derive from
(\ref{EofM1all}), (\ref{EofM2all}) the equations of
motion corresponding to the physical field of tensor
rank-$n$
\begin{eqnarray}
&&
\Delta{}Q|\chi^{0}_{0}\rangle_n
+\frac{1}{2}\bigl\{\tilde{T}_0,q_1^+q_1\bigr\}
|\chi^{1}_{0}\rangle_n
=0,
\label{EofM1}
\\&&
\tilde{T}_0|\chi^{0}_{0}\rangle_n +
\Delta{}Q|\chi^{1}_{0}\rangle_n =0,
\label{EofM2}
\end{eqnarray}
where the $|\chi_0^0\rangle_n$, $|\chi_0^1\rangle_n$
are assumed to obey the relations
\begin{eqnarray}\label{schi}
\sigma|\chi^0_0\rangle_n =\bigl(n-d/2\bigr)|\chi^0_0\rangle_n,
&\qquad&
\sigma|\chi^1_0\rangle_n
=\bigl(n-d/2\bigr)|\chi^1_0\rangle_n
.
\end{eqnarray}

The field equations (\ref{EofM1}), (\ref{EofM2}) are Lagrangian
ones and can be deduced from the following Lagrangian\footnote{The
Lagrangian is defined as usual up to an overall factor.}
\begin{eqnarray}
{\cal{}L}_n
&=&
{}_n\langle\tilde{\chi}^{0}_{0}|K_n\tilde{T}_0|\chi^{0}_{0}\rangle_n
+
\frac{1}{2}\,{}_n\langle\tilde{\chi}^{1}_{0}|K_n\bigl\{
   \tilde{T}_0,q_1^+q_1\bigr\}|\chi^{1}_{0}\rangle_n
\nonumber
\\&&\qquad{}
+
{}_n\langle\tilde{\chi}^{0}_{0}|K_n\Delta{}Q|\chi^{1}_{0}\rangle_n
+
{}_n\langle\tilde{\chi}^{1}_{0}|K_n\Delta{}Q|\chi^{0}_{0}\rangle_n
,
\label{L1}
\end{eqnarray}
where the standard scalar product for the creation and
annihilation operators is assumed, and the operator $K_n$ is the
operator $K$ (\ref{K}), where the following substitution is made
$h\to{}d/2-n$.

The equations of motion (\ref{EofM1}), (\ref{EofM2}) and the
action (\ref{L1}) are invariant with respect to the gauge
transformations
\begin{eqnarray}
\delta|\chi^{0}_{0}\rangle_n
&=&
\Delta{}Q|\Lambda^{0}_{0}\rangle_n
 +
 \frac{1}{2}\bigl\{\tilde{T}_0,q_1^+q_1\bigr\}
 |\Lambda^{1}_{0}\rangle_n,
\label{GT1}
\\
\delta|\chi^{1}_{0}\rangle_n
&=&
\tilde{T}_0|\Lambda^{0}_{0}\rangle_n
 +\Delta{}Q|\Lambda^{1}_{0}\rangle_n
 ,
\label{GT2}
\end{eqnarray}
which are reducible, with the gauge parameters
$|\Lambda^{(i)}{}^{j}_{0}\rangle_n$, $j=0,1$
subject to the same conditions
as those for $|\chi^j_0\rangle_n$ in (\ref{schi}),
\begin{align}
\delta|\Lambda^{(i)}{}^{0}_{0}\rangle_n
&=
\Delta{}Q|\Lambda^{(i+1)}{}^{0}_{0}\rangle_n
 +
 \frac{1}{2}\bigl\{\tilde{T}_0,q_1^+q_1\bigr\}
 |\Lambda^{(i+1)}{}^{1}_{0}\rangle_n,
&
|\Lambda^{(0)}{}^0_0\rangle_n=|\Lambda^0_0\rangle_n,
\label{GTi1}
\\
\delta|\Lambda^{(i)}{}^{1}_{0}\rangle_n
&=
\tilde{T}_0|\Lambda^{(i+1)}{}^{0}_{0}\rangle_n
 +\Delta{}Q|\Lambda^{(i+1)}{}^{1}_{0}\rangle_n,
&
|\Lambda^{(0)}{}^1_0\rangle_n=|\Lambda^1_0\rangle_n,
\label{GTi2}
\end{align}
with finite number of reducibility stages $i_{max}=n-1$.

We now determine the value of the arbitrary parameter $m_0$ using
the condition that the equations (\ref{TheEM}) [or in operator form
(\ref{TheEM-op})] for the basic vector $|\psi\rangle$
(\ref{PhysState}) be reproduced. To this end it is necessary that
conditions (\ref{TheEM}) be implied by Eqs. (\ref{EofM1}),
(\ref{EofM2}). Note that the general vector $|\chi^0_0\rangle_n$
includes the basic vector $|\psi\rangle$ (\ref{PhysState})
\begin{equation}
\label{relation}
|\chi^0_0\rangle_n
=
|\psi\rangle_n + |\psi_A\rangle_n,
\qquad
\left.\phantom{\Bigl[}|\psi_A\rangle_n\right|_{ghosts=b^+=f^+=0}=0
.
\end{equation}
In appendix~\ref{reduction} we shall demonstrate that due to the gauge
fixing
and a part of the equations of motion the vector $|\psi_A\rangle_n$ can be
completely removed and the resulting equations of motion have the form
\begin{equation}
\label{result eqs}
T_0|\psi\rangle_n  =
(t_0-\tilde{\gamma}m_0)|\psi\rangle_n =0,
\quad
T_1|\psi\rangle_n =
t_1|\psi\rangle_n =0,
\quad
L_1|\psi\rangle_n=
l_1|\psi\rangle_n=0,
\end{equation}
so the action actually
reproduces the correct equations of motion (\ref{TheEM}).
The above relations permit one to determine the parameter $m_0$ in a
unique way as follows
\begin{equation}
\label{final m0}
m_0
= m +r^{\frac{1}{2}}\bigl(d/2-n\bigr)
=m+r^{\frac{1}{2}}h
.
\end{equation}
Thus we
have constructed Lagrangians for antisymmetric fermionic fields of any
tensor rank using the BRST approach.

Finally we note that Lagrangian (\ref{L1}) can be simplified.
In particular one can remove all the auxiliary field and write Lagrangian
$\mathcal{L}_n$ for rank-$n$ antisymmetric fermionic field in
terms of basic field $\psi_{\mu_1\ldots\mu_n}$ only\footnote{Lagrangian
(\ref{kirdyk}) is Lagrangian (\ref{L1}) multiplied by $(-1)^n$. See
(\ref{kirdyk'}).}
(see details in appendix~\ref{simple})
\begin{eqnarray}
\mathcal{L}_n&=&
\sum_{k=0}^{n}
\frac{1}{(n-k)!}\;\bar{\psi}^{\mu_1\ldots\mu_{n-k}}
[(-1)^{k}i\gamma^\sigma\nabla_\sigma-m_0]\psi_{\mu_1\ldots\mu_{n-k}}
\nonumber
\\
&&{}
-i\sum_{k=0}^{n-1}\frac{(-1)^{k}}{(n-k-1)!}\;
\Bigl(
\bar{\psi}^{\mu_1\ldots\mu_{n-k}}\nabla_{\mu_1}\psi_{\mu_2\ldots\mu_{n-k}}
+
\bar{\psi}^{\mu_2\ldots\mu_{n-k}}\nabla^{\mu_1}\psi_{\mu_1\ldots\mu_{n-k}}
\Bigr)
,
\label{kirdyk}
\end{eqnarray}
where $m_0=m+r^{\frac{1}{2}}(d/2-n)$ with $m$ being the mass and we have denoted
\begin{eqnarray}
\psi_{\mu_{k+1}\ldots\mu_n}=
\frac{1}{k!}\gamma^{\mu_k}\ldots\gamma^{\mu_1}\psi_{\mu_1\ldots\mu_n}
,
&\qquad&
\bar{\psi}_{\mu_{k+1}\ldots\mu_n}=
\frac{1}{k!}
\bar{\psi}_{\mu_1\ldots\mu_n}\gamma^{\mu_1}\ldots\gamma^{\mu_k}
.
\end{eqnarray}
If we put the mass $m=0$ in (\ref{kirdyk}) (which means
$m_0=r^{\frac{1}{2}}(d/2-n)$)  then one should expect that
Lagrangian (\ref{kirdyk}) becomes gauge invariant Lagrangian for the
rank-$n$ massless antisymmetric fermionic field.

\section{Summary}\label{secSummary}
We have constructed the Lagrangian formulation for massive fermionic
antisymmetric tensor field theory in $AdS_d$ space and found the
various equivalent forms of the Lagrangian. In general, the
Lagrangian contains basic field together with a number of auxiliary
and St\"uckelberg fields determining the reducible gauge model. Such
a situation is a standard for massive higher spin field theories.
However, the specific features namely fermionic antisymmetric field
allowed to eliminate completely all the auxiliary and St\"uckelberg
fields from the action and obtain the Lagrangian only in terms of
basic field. As far as we know, such a Lagrangian has never been
presented before in the literature\footnote{One points out that the
standard (local) Lagrangian of totally symmetric massive higher spin
field theory with auxiliary fields can be transformed to equivalent
but nonlocal Lagrangian without auxiliary fields \cite{Francia}. In
the given paper, speaking about Lagrangian without auxiliary fields
we mean a local Lagrangian without auxiliary fields. Also we
emphasize that the Lagrangian formulations for totally antisymmetric
tensor-spinor fields, which were studied in the given paper, have
never been constructed earlier at all.}.

We have demonstrated that if we don't include in the equations of
motion for antisymmetric field the terms with the inverse powers of
the mass then the equations of motion in curved space of arbitrary
dimension are consistent only in space of constant curvature. Then
we have shown that the BRST approach which was earlier applied for
totally symmetric or mixed symmetry higher spin fields perfectly
works for massive fermionic antisymmetric fields in $AdS_d$ space.

The initial point of Lagrangian construction is reformulating the
massive irreducible representation of the $AdS_d$ on fermionic
antisymmetric tensor fields as operator constraints in auxiliary
Fock space. Then we found the closed algebra generated by these
operators and applied the BRST construction \cite{0608005}. As a
result we obtained the reducible gauge Lagrangian theory, the
corresponding Lagrangian and (St\"uckelberg) gauge transformations
are given by (\ref{L1}), (\ref{GT1})--(\ref{GTi2}) and the order of
reducibility grows with the value of the rank of the antisymmetric
field. Like all the Lagrangians constructed on the base of the BRST
approach, the Lagrangian in the case under consideration possess
rich gauge symmetry and contain many auxiliary fields. Partially
fixing some of the symmetries or/and eliminating some auxiliary
fields it is possible to derive the various intermediate Lagrangian
formulations. In particular one can write Lagrangian with some
number of auxiliary fields without gauge symmetry (\ref{L1-5}) or
with gauge symmetry (\ref{L1-4}). In particular, the Lagrangian in
terms of basic field only (i.e. without any auxiliary fields and
gauge symmetries) is also obtained (\ref{kirdyk}).

\section*{Acknowledgements}
The authors are grateful to R.R.~Metsaev and Yu.M.~Zinoviev for valuable comments.
The work of I.L.B, V.A.K and  L.L.R. was partially supported by RF
Presidential grant for LSS, project No.\ 4489.2006.2 and by the RFBR
grant, project No.\ 09-02-00078-a. The work of I.L.B and V.A.K was
partially supported by the INTAS grant, project INTAS-05-7928.
I.L.B. is grateful to joint RFBR-Ukraine grant, project No.\
08-02-90490.

\appendix

\section{Calculation of the additional parts}\label{AdParts}
\renewcommand{\theequation}{\Alph{section}.\arabic{equation}}
\setcounter{equation}{0}

In this Appendix we show how the representation of the algebra given
in Table~\ref{table'} can be constructed in terms of some creation
and annihilation operators.

Let us consider a representation of this algebra with the vector
$|0\rangle_V$ annihilated by the operators $l_1'$ and
$t_1'$
\begin{eqnarray}
\label{vac}
l_1'|0\rangle_V=t_1'|0\rangle_V=0,
\end{eqnarray}
and being the eigenvector of the operators $t_0'$, $l_0'$,
$g_0'$ and $g_m'$
\begin{eqnarray}
\label{Her-vac}
&&
t_0'|0\rangle_V=-\tilde{\gamma}m_0|0\rangle_V,
\quad
l_0'|0\rangle_V=-m_0^2|0\rangle_V,
\quad
g_0'|0\rangle_V=h|0\rangle_V,
\quad
g_m'|0\rangle_V=h_m|0\rangle_V,
\end{eqnarray}
where $m_0$, $h_m$ are arbitrary constants with dimension of
mass and $h$ is an arbitrary dimensionless constant.
They are the arbitrary constants which must be contained in the
additional parts of Hermitian operators.
Next we choose the basis vectors of this representation as follows
\begin{eqnarray}
\label{BVectors}
|0,n\rangle_V=(t_1^{\prime+})^n|0\rangle_V,
&\qquad&
|1,n\rangle_V=\frac{l_1^{\prime+}}{m_1}(t_1^{\prime+})^n|0\rangle_V,
\end{eqnarray}
where $m_1$ is an arbitrary nonzero constant with dimension of
mass. It may be constructed from the parameters of the theory
$m_1=f(m,r)\neq0$.

Now using commutators given in Table~\ref{table'} and (\ref{vac})--(\ref{BVectors}) one finds
\begin{eqnarray}
\label{A4}
t_1^{\prime+}|0,n\rangle_V=|0,n+1\rangle_V,
&\qquad&
t_1^{\prime+}|1,n\rangle_V=|1,n+1\rangle_V,
\\
l_1^{\prime+}|0,n\rangle_V=m_1|1,n\rangle_V,
&\qquad&
l_1^{\prime+}|1,n\rangle_V=-\frac{r}{4m_1}\,|0,n+2\rangle_V,
\\
g_0'|0,n\rangle_V=(n+h)|0,n\rangle_V,
&\qquad&
g_0'|1,n\rangle_V=(n+1+h)|1,n\rangle_V,
\\
g_m'|0,n\rangle_V=h_m|0,n\rangle_V,
&\qquad&
g_m'|1,n\rangle_V=h_m|1,n\rangle_V,
\end{eqnarray}
\begin{eqnarray}
&&
t_0'|0,n\rangle_V=
-2nm_1|1,n-1\rangle_V-\tilde{\gamma}m_0|0,n\rangle_V,
\\
&&
t_0'|1,n\rangle_V=
\frac{r}{2m_1}\Bigl(
n+1+2h
\Bigr)|0,n+1\rangle_V
-\tilde{\gamma}m_0|1,n\rangle_V,
\\
&&
l_0'|0,n\rangle_V=-\bigl[rn(n+2h)+m_0^2\bigr]|0,n\rangle_V,
\\
&&
l_0'|1,n\rangle_V=-\bigl[rn(n+2h)+m_0^2\bigr]|1,n\rangle_V
   -2r(n+h+{\textstyle\frac{1}{2}})|1,n\rangle_V,
\\
&&
t_1'|0,n\rangle_V=n(n-1+2h)|0,n-1\rangle_V,
\\
&&
t_1'|1,n\rangle_V=
\tilde{\gamma}\frac{m_0}{m_1}|0,n\rangle_V
+n(n+1+2h)|1,n-1\rangle_V
\\
&&
l_1'|0,n\rangle_V=n\tilde{\gamma}m_0|0,n-1\rangle_V
+n(n-1)m_1|1,n-2\rangle_V
\\
&&
l_1'|1,n\rangle_V=n\tilde{\gamma}m_0|1,n-1\rangle_V
-\frac{r}{4m_1}n(n-1)|0,n\rangle_V
\nonumber
\\
&&\hspace*{10em}{}
+\frac{m_0^2-r(h+\frac{1}{2})(n+h)}{m_1}|0,n\rangle_V
.
\label{A15}
\end{eqnarray}

Now let us turn to construction of a representation of the
operator algebra given in Table~\ref{table'} in terms of
creation and annihilation operators.
The number of pairs of these operators and their statistics is
defined by the number and the statistics of the operators used in the definition of the
basis vectors (\ref{BVectors}).
Thus we introduce one pair of the bosonic and one pair of fermionic creation and
annihilation operators with the standard commutation relations
\begin{eqnarray}
[b,b^+]=1,
\qquad
\{f,f^+\}=1,
\end{eqnarray}
corresponding to $t_1'$, $t_1^{\prime+}$ and $l_1'$,
$l_1^{\prime+}$ respectively.
After this we map the basis vectors (\ref{BVectors}) and the basis
vectors of the Fock space generated by $b^+$, $f^+$
\begin{eqnarray}
|0,n\rangle_V&\longleftrightarrow&
(b^+)^{n}|0\rangle=|0,n\rangle=|n\rangle,
\\
|1,n\rangle_V&\longleftrightarrow&
f^+(b^+)^{n}|0\rangle=|1,n\rangle,
\end{eqnarray}
and find from (\ref{A4})--(\ref{A15}) form of the operators in terms of the creation and
annihilation operators $b$, $b^+$, $f$, $f^+$.
They are given by relations (\ref{ad-1})--(\ref{ad-5}).

\section{Reduction to the initial irreducible relations}\label{reduction}

Let us show that the equations of motion (\ref{TheEM}),
[or equivalently in operatorial form (\ref{TheEM-op})] can be obtained from
the Lagrangian (\ref{L1}) after gauge-fixing and removing the
auxiliary fields by using a part of the equations of motion. Let
us start with gauge-fixing.

\subsection{Gauge-fixing}

Let us consider antisymmetric fermionic field of tensor rank $n$.
Then we have a
reducible gauge theory with $n-1$ reducibility stages. Due to
restriction (\ref{schi}) and the ghost number restriction [see the
right-hand formulae in (\ref{0L})],  the lowest-stage gauge
parameters have the form
\begin{eqnarray}
\label{gauge param}
|\Lambda^{(n-1)}{}_0^0\rangle_n
&=&
(p^+_1)^{n-1}\left\{\mathcal{P}_1^+|\lambda \rangle_0 +
{p}_1^+|\lambda_1
\rangle_0\right\}, \\
|\Lambda^{(n-1)}{}_0^1\rangle_n
& \equiv &0,
\end{eqnarray}
with the subscripts of the state vectors being
associated with the eigenvalues of the corresponding state vectors
(\ref{schi}).
In what follows we shall omit these subscripts.
We see that gauge parameter
$|\Lambda^{(n-1)}{}_0^0\rangle$ cannot depend (in particular) on
$f^+$.
It can be verified directly that
one can eliminate
the dependence on $f^+$ from the gauge
function $|\Lambda^{(n-2)}{}^0_0\rangle$ of the $(n-2)$-th
stage.
The gauge function $|\Lambda^{(n-2)}{}^1_0\rangle$ has no
$f^+$ dependence due to the same reason as
$|\Lambda^{(n-1)}{}_0^0\rangle$.
It is then possible to verify that one can remove the dependence of
$|\Lambda^{(n-3)}{}^0_0\rangle$, $|\Lambda^{(n-3)}{}^1_0\rangle$
on $f^+$ with the
help of the remaining gauge parameters
$|\Lambda^{(n-2)}{}^0_0\rangle$,
$|\Lambda^{(n-2)}{}^1_0\rangle$ which do not depend on
$f^+$.

We now suppose that we have removed the dependence on
$f^+$ from the gauge functions of the $i$-th stage
$|\Lambda^{(i)}{}^j_0\rangle$, $j=0,1$, i.e., we have
$f|\Lambda^{(i)}{}^j_0\rangle=0$.
Let us consider the gauge transformation for
$|\Lambda^{(i-1)}{}^j_0\rangle$.
It has the following structure
\begin{eqnarray}
\delta|\Lambda^{(i-1)}{}^j_0\rangle
&=&
q_1L_1^+|\Lambda^{(i)}{}^j_0\rangle+\ldots
=
m_1q_1f^+|\Lambda^{(i)}{}^j_0\rangle+\ldots
\end{eqnarray}
The $f^+$ dependent part of $|\Lambda^{(i-1)}{}^j_0\rangle$ and
$q_1f^+|\Lambda^{(i)}{}^j_0\rangle$ have the same decomposition
on creation operators (\ref{chi}). Therefore we can eliminate
the $f^+$ dependent part of $|\Lambda^{(i-1)}{}^j_0\rangle$
having used all the restricted gauge parameters
$|\Lambda^{(i)}{}^j_0\rangle$.

The same argumentation is valid for the gauge transformations of
fields $|\chi_0^j\rangle$.
But in this case we do not use all the gauge parameters since in
$|\Lambda_0^0\rangle$ there are terms independent of $p_1^+$ and
they are annihilated by $q_1f^+$.
These terms have the following ghost structure
\begin{eqnarray}
|\Lambda_0^0\rangle&=&\mathcal{P}_1^+|\lambda\rangle+\ldots
\end{eqnarray}
where $|\lambda\rangle$ depends on $a^{+\mu}$ and $b^+$ and
independent of $f^+$ due to
the condition $f|\Lambda_0^0\rangle=0$.
We can use the remaining gauge parameter $|\lambda\rangle$ to
eliminate the $b^+$ dependence in the ghost independent part of
$|\chi_0^0\rangle$ with the help of the transformation
\begin{eqnarray}
\delta|\chi_0^0\rangle&=&\eta_1T_1^+\;\mathcal{P}_1^+|\lambda\rangle+\ldots
=
b^+|\lambda\rangle+\ldots
\end{eqnarray}
Now we have used all the gauge parameters.
Thus the gauge conditions on the fields are
\begin{eqnarray}
f|\chi_0^0\rangle=f|\chi_0^1\rangle=0,
&\qquad&
p_1\mathcal{P}_1b|\chi_0^0\rangle=0.
\label{GT}
\end{eqnarray}

Let us turn to the elimination of the rest auxiliary fields with
the help of the equations of motion.

\subsection{Removing of the auxiliary fields with the equations
of motion}

Let us decompose the equations of motion (\ref{EofM1}),
(\ref{EofM2}) on $f^+$. The equations of motion at $f^+$ are
\begin{eqnarray}
(q_1^+b^2+q_1)|\chi_0^1\rangle=2b|\chi_0^0\rangle,
&\qquad&
(q_1^+b^2+q_1)|\chi_0^0\rangle=2q_1^+q_1b|\chi_0^1\rangle.
\label{eq-f+}
\end{eqnarray}
Then we decomose fields $|\chi_0^0\rangle$ and $|\chi_0^1\rangle$
in ghosts $\eta_1^+$, $\mathcal{P}_1^+$
\begin{eqnarray}
\label{dec-0}
|\chi_0^j\rangle&=&
|\chi_{00}^{j0}\rangle
+\eta_1^+|\chi_{00}^{j1}\rangle
+\mathcal{P}_1^+|\chi_{01}^{j0}\rangle
+\eta_1^+\mathcal{P}_1^+|\chi_{01}^{j1}\rangle,
\qquad
j=0,1
\end{eqnarray}
and substitute this decomposition into (\ref{eq-f+}).
First we consider the following pair of equations corresponding to
$(\eta_1^+)^0(\mathcal{P}_1^+)^0$
\begin{eqnarray}
(q_1^+b^2+q_1)|\chi_{00}^{10}\rangle=2b|\chi_{00}^{00}\rangle,
&\qquad&
(q_1^+b^2+q_1)|\chi_{00}^{00}\rangle=2q_1^+q_1b|\chi_{00}^{10}\rangle.
\label{eq-f0}
\end{eqnarray}
Decomposing fields $|\chi_{00}^{j0}\rangle$ in power series
of bosonic ghosts $q_1^+$, $p_1^+$
\begin{eqnarray}
|\chi_{00}^{00}\rangle=\sum_{k=0}^{[n/2]}
\frac{(-iq_1^+p_1^+)^k}{k!}|\chi_{00k}^{00}\rangle,
&\qquad&
|\chi_{00}^{10}\rangle=\sum_{k=1}^{[(n+1)/2]}(q_1^+)^{k-1}
\frac{(-ip_1^+)^k}{k!}|\chi_{00k}^{10}\rangle
\label{dec-1}
\end{eqnarray}
where fields $|\chi_{00k}^{j0}\rangle$ have ghost number equal
to zero $gh(|\chi_{00k}^{j0}\rangle)=0$.
Substituting (\ref{dec-1}) into (\ref{eq-f0})
and considering the obtained equations from the lowest power of
$p_1^+$ (and taking into account the gauge
$b|\chi_{000}^{00}\rangle=0$)
we get that all $|\chi_{00k}^{j0}\rangle=0$, $k\ge1$.
That is we have $|\chi_{00}^{10}\rangle=0$ and
$|\chi_{00}^{00}\rangle=|\psi\rangle$, with $|\psi\rangle$ being
the physical field (\ref{PhysState}).

Next we consider one more pair of equations (\ref{eq-f+})
corresponding to $(\eta_1^+)^1(\mathcal{P}_1^+)^0$ coefficient
of decomposition (\ref{dec-0})
\begin{eqnarray}
(q_1^+b^2+q_1)|\chi_{00}^{11}\rangle=2b|\chi_{00}^{01}\rangle,
&\qquad&
(q_1^+b^2+q_1)|\chi_{00}^{01}\rangle=2q_1^+q_1b|\chi_{00}^{11}\rangle.
\label{eq-f1}
\end{eqnarray}
Doing decomposition of the fields in power series of ghosts
$q_1^+$, $p_1^+$ analogous to (\ref{dec-1}) and considering
equations from the lowest powers of $p_1^+$ one concludes that
$|\chi_{00}^{01}\rangle=|\chi_{00}^{11}\rangle=0$.

Let us now turn to the equations which are coefficients of
equations (\ref{EofM1}), (\ref{EofM2}) at
$(f^+)^0(\eta_1^+)^0(\mathcal{P}_1^+)^0$
\begin{eqnarray}
\label{100}
&&
T_0|\psi\rangle-2ip_1^+|\chi_{01}^{00}\rangle+T_1^+|\chi_{01}^{10}\rangle=0,
\\
&&
q_1^+L_1|\psi\rangle+T_1^+|\chi_{01}^{00}\rangle
+q_1^+(1-2ip_1^+q_1)|\chi_{01}^{10}\rangle=0
\label{101}
\end{eqnarray}
and
at $(f^+)^0(\eta_1^+)^1(\mathcal{P}_1^+)^0$
\begin{eqnarray}
\label{102}
&&
2ip_1^+|\chi_{01}^{01}\rangle-T_1^+|\chi_{01}^{11}\rangle=0,
\\
&&
T_1|\psi\rangle-T_1^+|\chi_{01}^{01}\rangle
-q_1^+(1-2ip_1^+q_1)|\chi_{01}^{11}\rangle=0,
\label{103}
\end{eqnarray}
where we have taken into account that
$|\chi_{00}^{00}\rangle=|\psi\rangle$ and
$|\chi_{00}^{10}\rangle=|\chi_{00}^{01}\rangle=|\chi_{00}^{11}\rangle=0$.

Let us consider the first pair of the equations. We decompose
fields $|\chi_{01}^{j0}\rangle$ in bosonic ghosts $q_1^+$, $p_1^+$
\begin{eqnarray}
|\chi_{01}^{00}\rangle
=q_1^+\sum_{k=0}^{[(n-2)/2]}\frac{(-iq_1^+p_1^+)^k}{k!}|\chi_{01k}^{00}\rangle,
&\quad&
|\chi^{10}_{01}\rangle
=\sum_{k=0}^{[(n-1)/2]}\frac{(-iq_1^+p_1^+)^k}{k!}|\chi_{01k}^{10}\rangle,
\end{eqnarray}
where all fields $|\chi_{01k}^{j0}\rangle$ have ghost number
equal to zero.
Starting from the highest power of $p_1^+$ we conclude that all
the $|\chi_{01k}^{j0}\rangle=0$ except
$|\chi_{010}^{10}\rangle$. Now equation (\ref{100}) reduce to
\begin{eqnarray}
\label{105}
T_0|\psi\rangle+T_1^+|\chi_{010}^{10}\rangle=0.
\end{eqnarray}
Decomposing field $|\chi_{010}^{10}\rangle$ in power series of
creation operator $b^+$ and substituting this decomposition into
(\ref{105}) we find that as a result $|\chi_{010}^{10}\rangle=0$.
That is we get
$|\chi_{01}^{00}\rangle=|\chi_{01}^{10}\rangle=0$.

Similar consideration of equations (\ref{102}), (\ref{103})
leads us to conclusion that
$|\chi_{01}^{01}\rangle=|\chi_{01}^{11}\rangle=0$.

Thus we have shown that all the auxiliary fields are equal to
zero due to the gauge condition (\ref{GT}) or as a solution to the
equations of motion.
The equations of motion on the physical field $|\psi\rangle$
followed from (\ref{100}), (\ref{103}), (\ref{101}) are
\begin{eqnarray}
T_0|\psi\rangle=(t_0-\tilde{\gamma}m_0)|\psi\rangle=0,
\quad
T_1|\psi\rangle=t_1|\psi\rangle=0,
\quad
L_1|\psi\rangle=l_1|\psi\rangle=0
\end{eqnarray}
which coincide with (\ref{result eqs}) and with (\ref{TheEM-op})
or in component form with (\ref{TheEM}).

\section{Simplified Lagrangians}\label{simple}
\renewcommand{\theequation}{\Alph{section}.\arabic{equation}}
\setcounter{equation}{0}

Let us try to simplify Lagrangian (\ref{L1}) and write it in terms of the
physical field only.
For this purpose we decompose fields $|\chi_{0}^{0}\rangle$,
$|\chi^{1}_{0}\rangle$ in power series of fermionic ghost fields $\eta_1^+$,
$\mathcal{P}_1^+$ (\ref{dec-0}) and substitute into (\ref{L1}).
One has
\begin{eqnarray}
{\cal{}L}_n
&=&
\langle\chi_{00}^{00}|K_n\Bigl\{
T_0|\chi_{00}^{00}\rangle
-2ip_1^+|\chi_{01}^{00}\rangle
+\frac{r}{2}(G_0-2g_0')q_1^+|\chi_{00}^{01}\rangle
\nonumber
\\
&&\qquad{}
+
(q_1^+L_1+q_1L_1^+)|\chi_{00}^{10}\rangle
+T_1^+|\chi_{01}^{10}\rangle
-\frac{r}{4}\Bigl[
q_1(T_1^+-2t_1^{\prime+})+q_1^+(T_1-2t_1')
\Bigr]q_1^+|\chi_{00}^{11}\rangle
\Bigr\}
\nonumber
\\
&&{}
+\langle\chi_{01}^{00}|K_n\Bigl\{
-T_0|\chi_{00}^{01}\rangle
+2ip_1|\chi_{00}^{00}\rangle
+2ip_1^+|\chi_{01}^{01}\rangle
+
T_1|\chi_{00}^{10}\rangle
-(q_1^+L_1+q_1L_1^+)|\chi_{00}^{11}\rangle
-T_1^+|\chi_{01}^{11}\rangle
\Bigr\}
\nonumber
\\
&&{}
+\langle\chi_{00}^{01}|K_n\Bigl\{
-T_0|\chi_{01}^{00}\rangle
+\frac{r}{2}(G_0-2g_0')q_1|\chi_{00}^{00}\rangle
+\frac{r}{2}(G_0-2g_0')q_1^+|\chi_{01}^{01}\rangle
\nonumber
\\
&&\qquad{}
-(q_1^+L_1+q_1L_1^+)|\chi_{01}^{10}\rangle
-\frac{r}{4}\Bigl[
q_1(T_1^+-2t_1^{\prime+})+q_1^+(T_1-2t_1')
\Bigr]
\Bigl(q_1|\chi_{00}^{10}\rangle+q_1^+|\chi_{01}^{11}\rangle\Bigr)
\Bigr\}
\nonumber
\\
&&{}
-\langle\chi_{01}^{01}|K_n\Bigl\{
T_0|\chi_{01}^{01}\rangle
+2ip_1|\chi_{01}^{00}\rangle
-\frac{r}{2}(G_0-2g_0')q_1|\chi_{00}^{01}\rangle
\nonumber
\\
&&\qquad{}
+(q_1^+L_1+q_1L_1^+)|\chi_{01}^{11}\rangle
+T_1|\chi_{01}^{10}\rangle
+\frac{r}{4}\Bigl[
q_1(T_1^+-2t_1^{\prime+})+q_1^+(T_1-2t_1')
\Bigr]
q_1|\chi_{00}^{11}\rangle
\Bigr\}
\nonumber
\\
&&{}
+\langle\chi_{00}^{10}|K_n\Bigl\{
T_0q_1^+q_1|\chi_{00}^{10}\rangle
-2iq_1^+p_1^+q_1|\chi_{01}^{10}\rangle
+q_1^+|\chi_{01}^{10}\rangle
+\frac{r}{2}(G_0-2g_0')q_1^{+2}q_1|\chi_{00}^{11}\rangle
\nonumber
\\
&&\qquad{}
+
(q_1^+L_1+q_1L_1^+)|\chi_{00}^{00}\rangle
+T_1^+|\chi_{01}^{00}\rangle
-\frac{r}{4}\Bigl[
q_1(T_1^+-2t_1^{\prime+})+q_1^+(T_1-2t_1')
\Bigr]q_1^+|\chi_{00}^{01}\rangle
\Bigr\}
\nonumber
\\
&&{}
+\langle\chi_{01}^{10}|K_n\Bigl\{
-T_0q_1^+q_1|\chi_{00}^{11}\rangle
+2iq_1^+q_1p_1|\chi_{00}^{10}\rangle
+q_1|\chi_{00}^{10}\rangle
-q_1^+|\chi_{01}^{11}\rangle
+2iq_1^+p_1^+q_1|\chi_{01}^{11}\rangle
\nonumber
\\
&&\qquad{}
+
T_1|\chi_{00}^{00}\rangle
-(q_1^+L_1+q_1L_1^+)|\chi_{00}^{01}\rangle
-T_1^+|\chi_{01}^{01}\rangle
\Bigr\}
\nonumber
\\
&&{}
+\langle\chi_{00}^{11}|K_n\Bigl\{
-T_0q_1^+q_1|\chi_{01}^{10}\rangle
+\frac{r}{2}(G_0-2g_0')q_1^+q_1
\Bigl(q_1|\chi_{00}^{10}\rangle+q_1^+|\chi_{01}^{11}\rangle\Bigr)
\nonumber
\\
&&\qquad{}
-(q_1^+L_1+q_1L_1^+)|\chi_{01}^{00}\rangle
-\frac{r}{4}\Bigl[
q_1(T_1^+-2t_1^{\prime+})+q_1^+(T_1-2t_1')
\Bigr]
\Bigl(q_1|\chi_{00}^{00}\rangle+q_1^+|\chi_{01}^{01}\rangle\Bigr)
\Bigr\}
\nonumber
\\
&&{}
-\langle\chi_{01}^{11}|K_n\Bigl\{
T_0q_1^+q_1|\chi_{01}^{11}\rangle
+2iq_1^+q_1p_1|\chi_{01}^{10}\rangle
+q_1|\chi_{01}^{10}\rangle
-\frac{r}{2}(G_0-2g_0')q_1^+q_1^2|\chi_{00}^{11}\rangle
\nonumber
\\
&&\qquad{}
+(q_1^+L_1+q_1L_1^+)|\chi_{01}^{01}\rangle
+T_1|\chi_{01}^{00}\rangle
+\frac{r}{4}\Bigl[
q_1(T_1^+-2t_1^{\prime+})+q_1^+(T_1-2t_1')
\Bigr]
q_1|\chi_{00}^{01}\rangle
\Bigr\}
.
\label{L1-comp}
\end{eqnarray}
Then we partially fix the gauge analogously to as in Appendix~\ref{reduction} so
that parameters $|\Lambda_0^0\rangle$ and $|\Lambda_0^1\rangle$ do not depend
on $f^+$: $f|\Lambda_0^0\rangle=f|\Lambda_0^1\rangle=0$.
After the partial gauge fixing the gauge transformations of the fields
(\ref{GT1}) and (\ref{GT2}) become irreducible.
Decomposed the gauge parameters
$|\Lambda_0^0\rangle$ and $|\Lambda_0^1\rangle$
analogously to (\ref{dec-0}) we substitute them into (\ref{GT1}) and
(\ref{GT2}). The result is
\begin{eqnarray}
\delta|\chi_{00}^{00}\rangle
&=&
(q_1^+L_1+q_1L_1^+)|\Lambda_{00}^{00}\rangle
+T_1^+|\Lambda_{01}^{00}\rangle
-\frac{r}{4}\Bigl[
q_1(T_1^+-2t_1^{\prime+})+q_1^+(T_1-2t_1')
\Bigr]q_1^+|\Lambda_{00}^{01}\rangle
\nonumber
\\
&&{}
+T_0q_1^+q_1|\Lambda_{00}^{10}\rangle
-2iq_1^+p_1^+q_1|\Lambda_{01}^{10}\rangle
+q_1^+|\Lambda_{01}^{10}\rangle
+\frac{r}{2}(G_0-2g_0')q_1^{+2}q_1|\Lambda_{00}^{11}\rangle
\label{GTc-1}
\\
\delta|\chi_{00}^{01}\rangle
&=&
T_1|\Lambda_{00}^{00}\rangle
-(q_1^+L_1+q_1L_1^+)|\Lambda_{00}^{01}\rangle
-T_1^+|\Lambda_{01}^{01}\rangle
\nonumber
\\
&&{}
-T_0q_1^+q_1|\Lambda_{00}^{11}\rangle
+2iq_1^+q_1p_1|\Lambda_{00}^{10}\rangle
+q_1|\Lambda_{00}^{10}\rangle
-q_1^+|\Lambda_{01}^{11}\rangle
+2iq_1^+p_1^+q_1|\Lambda_{01}^{11}\rangle
\\
\delta|\chi_{01}^{00}\rangle
&=&
-(q_1^+L_1+q_1L_1^+)|\Lambda_{01}^{00}\rangle
-\frac{r}{4}\Bigl[
q_1(T_1^+-2t_1^{\prime+})+q_1^+(T_1-2t_1')
\Bigr]
\Bigl(q_1|\Lambda_{00}^{00}\rangle+q_1^+|\Lambda_{01}^{01}\rangle\Bigr)
\nonumber
\\
&&{}
-T_0q_1^+q_1|\Lambda_{01}^{10}\rangle
+\frac{r}{2}(G_0-2g_0')q_1^+q_1
\Bigl(q_1|\Lambda_{00}^{10}\rangle+q_1^+|\Lambda_{01}^{11}\rangle\Bigr)
\\
\delta|\chi_{01}^{01}\rangle
&=&
(q_1^+L_1+q_1L_1^+)|\Lambda_{01}^{01}\rangle
+T_1|\Lambda_{01}^{00}\rangle
+\frac{r}{4}\Bigl[
q_1(T_1^+-2t_1^{\prime+})+q_1^+(T_1-2t_1')
\Bigr]
q_1|\Lambda_{00}^{01}\rangle
\nonumber
\\
&&{}
+T_0q_1^+q_1|\Lambda_{01}^{11}\rangle
+2iq_1^+q_1p_1|\Lambda_{01}^{10}\rangle
+q_1|\Lambda_{01}^{10}\rangle
-\frac{r}{2}(G_0-2g_0')q_1^+q_1^2|\Lambda_{00}^{11}\rangle
\\
\delta|\chi_{00}^{10}\rangle
&=&
T_0|\Lambda_{00}^{00}\rangle
-2ip_1^+|\Lambda_{01}^{00}\rangle
+\frac{r}{2}(G_0-2g_0')q_1^+|\Lambda_{00}^{01}\rangle
\nonumber
\\
&&{}
+(q_1^+L_1+q_1L_1^+)|\Lambda_{00}^{10}\rangle
+T_1^+|\Lambda_{01}^{10}\rangle
-\frac{r}{4}\Bigl[
q_1(T_1^+-2t_1^{\prime+})+q_1^+(T_1-2t_1')
\Bigr]q_1^+|\Lambda_{00}^{11}\rangle
\\
\delta|\chi_{00}^{11}\rangle
&=&
-T_0|\Lambda_{00}^{01}\rangle
+2ip_1|\Lambda_{00}^{00}\rangle
+2ip_1^+|\Lambda_{01}^{01}\rangle
+T_1|\Lambda_{00}^{10}\rangle
-(q_1^+L_1+q_1L_1^+)|\Lambda_{00}^{11}\rangle
-T_1^+|\Lambda_{01}^{11}\rangle
\\
\delta|\chi_{01}^{10}\rangle
&=&
-T_0|\Lambda_{01}^{00}\rangle
+\frac{r}{2}(G_0-2g_0')q_1|\Lambda_{00}^{00}\rangle
+\frac{r}{2}(G_0-2g_0')q_1^+|\Lambda_{01}^{01}\rangle
\nonumber
\\
&&{}
-(q_1^+L_1+q_1L_1^+)|\Lambda_{01}^{10}\rangle
-\frac{r}{4}\Bigl[
q_1(T_1^+-2t_1^{\prime+})+q_1^+(T_1-2t_1')
\Bigr]
\Bigl(q_1|\Lambda_{00}^{10}\rangle+q_1^+|\Lambda_{01}^{11}\rangle\Bigr)
\\
\delta|\chi_{01}^{11}\rangle
&=&
T_0|\Lambda_{01}^{01}\rangle
+2ip_1|\Lambda_{01}^{00}\rangle
-\frac{r}{2}(G_0-2g_0')q_1|\Lambda_{00}^{01}\rangle
\nonumber
\\
&&{}
+(q_1^+L_1+q_1L_1^+)|\Lambda_{01}^{11}\rangle
+T_1|\Lambda_{01}^{10}\rangle
+\frac{r}{4}\Bigl[
q_1(T_1^+-2t_1^{\prime+})+q_1^+(T_1-2t_1')
\Bigr]
q_1|\Lambda_{00}^{11}\rangle
\label{GTc-8}
\end{eqnarray}

Let us proceed the gauge fixing. Now we remove the fields depending
on ghost $\eta_1^+$. That is we get rid of fields
$|\chi_{00}^{01}\rangle$ (using $|\Lambda_{00}^{01}\rangle$ and
$|\Lambda_{00}^{10}\rangle$ completely), $|\chi^{01}_{01}\rangle$
(using $|\Lambda_{01}^{01}\rangle$ and $|\Lambda_{01}^{10}\rangle$
completely), $|\chi_{00}^{11}\rangle$ (using
$|\Lambda_{00}^{00}\rangle$ partially and
$|\Lambda_{00}^{11}\rangle$ completely), $|\chi_{01}^{11}\rangle$
(using $|\Lambda_{01}^{00}\rangle$ partially and
$|\Lambda_{01}^{11}\rangle$ completely). A part of parameters
$|\Lambda_{00}^{00}\rangle$ and $|\Lambda_{01}^{00}\rangle$ remains
unused. These unused gauge parameters we denote as
$|\Lambda_{000}^{00}\rangle$ and $|\Lambda_{010}^{00}\rangle$ and
they are defined from the following decomposition of
$|\Lambda_{00}^{00}\rangle$ and $|\Lambda_{01}^{00}\rangle$ in power
series of bosonic ghosts $q_1^+$, $p_1^+$
\begin{eqnarray}
|\Lambda_{00}^{00}\rangle
=
\sum_{k=0}^{[(n-1)/2]}(q_1^+)^k\frac{(-ip_1^+)^{k+1}}{(k+1)!}
|\Lambda_{00k}^{00}\rangle
.
&\qquad&
|\Lambda_{01}^{00}\rangle
=
\sum_{k=0}^{[(n-1)/2]}\frac{(-iq_1^+p_1^+)^k}{k!}
|\Lambda_{01k}^{00}\rangle
.
\end{eqnarray}
Here $|\Lambda_{00k}^{00}\rangle$, $|\Lambda_{01k}^{00}\rangle$ depend on
$a^{\mu+}$, $b^+$ only (they are independent of $f^+$ due to the gauge fixing
$f|\Lambda_0^j\rangle=0$).
After the last partial gauge fixing Lagrangian (\ref{L1-comp})
for the residuary
fields are
\begin{eqnarray}
{\cal{}L}_n
&=&
\langle\chi_{00}^{00}|K_n\Bigl\{
T_0|\chi_{00}^{00}\rangle
-2ip_1^+|\chi_{01}^{00}\rangle
+(q_1^+L_1+q_1L_1^+)|\chi_{00}^{10}\rangle
+T_1^+|\chi_{01}^{10}\rangle
\Bigr\}
\nonumber
\\
&&
+\langle\chi_{00}^{10}|K_n\Bigl\{
T_0q_1^+q_1|\chi_{00}^{10}\rangle
+q_1^+(1-2ip_1^+q_1)|\chi_{01}^{10}\rangle
+(q_1^+L_1+q_1L_1^+)|\chi_{00}^{00}\rangle
+T_1^+|\chi_{01}^{00}\rangle
\Bigr\}
\nonumber
\\
&&
+\langle\chi_{01}^{10}|K_n\Bigl\{
(1+2iq_1^+p_1)q_1|\chi_{00}^{10}\rangle
+T_1|\chi_{00}^{00}\rangle
\Bigr\}
+
\langle\chi_{01}^{00}|K_n\Bigl\{
2ip_1|\chi_{00}^{00}\rangle
+T_1|\chi_{00}^{10}\rangle
\Bigr\}
,
\label{L1-2}
\end{eqnarray}
%
%
%
%
%
%
%
%
%
%
%
%
%
Let us decompose the fields entering in Lagrangian (\ref{L1-2}) in power
series of bosonic ghosts $q_1^+$  $p_1^+$
\begin{align}
&
|\chi_{00}^{00}\rangle
=\sum_{k=0}^{[n/2]}\frac{(-iq_1^+p_1^+)^k}{k!}|\Psi_{n-2k}\rangle
,
&&
|\chi_{00}^{10}\rangle
=\sum_{k=0}^{[(n-1)/2]}(q_1^+)^k\frac{(-ip_1^+)^{k+1}}{(k+1)!}|\Psi_{n-2k-1}\rangle
,
\\
&
|\chi_{01}^{00}\rangle
=\sum_{k=0}^{[(n-2)/2]}(q_1^+)^{k+1}\frac{(-ip_1^+)^k}{k!}|A_{n-2k-2}\rangle
,
&&
|\chi_{01}^{10}\rangle
=
\sum_{k=0}^{[(n-1)/2]}\frac{(-iq_1^+p_1^+)^k}{k!}|A_{n-2k-1}\rangle
,
\end{align}
where all $|\Psi_k\rangle$ and $|A_k\rangle$ depend on $a^{\mu+}$,
$b^+$, $f^+$ only and their subindices coincide with the eigenvalues
of operator $\sigma$ (\ref{schi}). Substituting these decompositions
of the fields into (\ref{L1-2}) one finds
\begin{eqnarray}
\mathcal{L}_n&=&
\langle\tilde{\Psi}_n|K_n\Bigl\{
T_0|\Psi_n\rangle
+L_1^+|\Psi_{n-1}\rangle
+T_1^+|A_{n-1}\rangle
\Bigr\}
\nonumber
\\
&&{}
+\sum_{k=1}^{n-1}\langle\tilde{\Psi}_{k}|K_n\Bigl\{
T_0|\Psi_{k}\rangle
+L_1|\Psi_{k+1}\rangle
+L_1^+|\Psi_{k-1}\rangle
+(n-k)|A_{k}\rangle
+T_1^+|A_{k-1}\rangle
\Bigr\}
\nonumber
\\
&&{}
+\langle\tilde{\Psi}_0|K_n\Bigl\{
T_0|\Psi_0\rangle+L_1|\Psi_1\rangle+n|A_0\rangle
\Bigr\}
+
\sum_{k=0}^{n-1}\langle\tilde{A}_{k}|K_n\Bigl\{
(n-k)|\Psi_{k}\rangle+T_1|\Psi_{k+1}\rangle
\Bigr\}
\label{L1-4}
\end{eqnarray}
Solving the equation of motion of $\langle\tilde{A}|$ we can express all $|\Psi_k\rangle$ in terms of $|\Psi_n\rangle$
\begin{eqnarray}
\label{trace}
|\Psi_{n-k}\rangle
=\frac{(-1)^k}{k!}(T_1)^k|\Psi_n\rangle.
\end{eqnarray}
Now let us fix the gauge completely using the residual gauge parameters
$|\Lambda_{000}^{00}\rangle$ and $|\Lambda_{010}^{00}\rangle$.
With their help we get rid of the dependence of the field $|\Psi_n\rangle$ on
$f^+$ and $b^+$ respectively.
That is the gauge condition is
\begin{eqnarray}
f|\Psi_n\rangle=b|\Psi_n\rangle=0.
\label{gaugePsi}
\end{eqnarray}
Let us denote the part of fields $|\Psi_k\rangle$ and
$|A_k\rangle$ which are independent of $f^+$ and $b^+$ as $|\psi_k\rangle$ and
$|\alpha_k\rangle$ respecively.
Then due to (\ref{gaugePsi}) we have that $|\Psi_n\rangle=|\psi_n\rangle$ and
$|\psi_n\rangle$ is the physical field
and due to (\ref{trace}) we get that all other $|\Psi_k\rangle$ are also
independent of $f^+$, $b^+$
\begin{eqnarray}
|\Psi_{n-k}\rangle
=\frac{(-1)^k}{k!}(T_1)^k|\Psi_n\rangle
=\frac{(-1)^k}{k!}(T_1)^k|\psi_n\rangle
=\frac{(-1)^k}{k!}(t_1)^k|\psi_n\rangle
=|\psi_{n-k}\rangle
.
\label{tr}
\end{eqnarray}
Thus after the gauge fixing (\ref{gaugePsi}) Lagrangian (\ref{L1-4}) become
\begin{eqnarray}
\mathcal{L}_n&=&
\langle\tilde{\psi}_n|\Bigl\{
(t_0-\tilde{\gamma}m_0)|\psi_n\rangle
+l_1^+|\psi_{n-1}\rangle
+t_1^+|\alpha_{n-1}\rangle
\Bigr\}
\nonumber
\\
&&{}
+\sum_{k=1}^{n-1}\langle\tilde{\psi}_{n-k}|\Bigl\{
(t_0-\tilde{\gamma}m_0)|\psi_{n-k}\rangle
+l_1|\psi_{n-k+1}\rangle
+l_1^+|\psi_{n-k-1}\rangle
+k|\alpha_{n-k}\rangle
+t_1^+|\alpha_{n-k-1}\rangle
\Bigr\}
\nonumber
\\
&&{}
+\langle\tilde{\psi}_0|\Bigl\{
(t_0-\tilde{\gamma}m_0)|\psi_0\rangle
+l_1|\psi_1\rangle+n|\alpha_0\rangle
\Bigr\}
+
\sum_{k=1}^{n}\langle\tilde{\alpha}_{n-k}|\Bigl\{
k|\psi_{n-k}\rangle+t_1|\psi_{n-k+1}\rangle
\Bigr\}
\label{L1-5}
\end{eqnarray}
and it has no gauge symmetry.
Finally we can express all $|\psi_k\rangle$ through $|\psi_n\rangle$ using
(\ref{tr}) and write Lagrangian in terms of the physical field only
\begin{eqnarray}
\mathcal{L}_n&=&
\sum_{k=0}^{n}\frac{1}{(k!)^2}\;
\langle\tilde{\psi}_n|(t_1^+)^k
(t_0-\tilde{\gamma}m_0)(t_1)^k|\psi_n\rangle
\nonumber
\\
&&{}
-\sum_{k=0}^{n-1}\frac{1}{k!(k+1)!}\;
\langle\tilde{\psi}_n|(t_1^+)^k(l_1^+t_1+t_1^+l_1)(t_1)^k|\psi_n\rangle
.
\label{L1-fin}
\end{eqnarray}
Let us rewrite Lagrangian (\ref{L1-fin}) in the component form.
Using the explicit expressions of the operators and
\begin{eqnarray}
|\psi_n\rangle=\frac{(-i)^n}{n!}a^{+\mu_1}\ldots
a^{+\mu_n}\psi(x)_{\mu_1\ldots\mu_n}|0\rangle
&\qquad&
\langle\tilde{\psi}_n|=\langle0|\psi^+(x)_{\mu_1\ldots\mu_n}\tilde{\gamma}^0
a^{\mu_n}\ldots a^{\mu_1}\frac{i^n}{n!}
\end{eqnarray}
we find
\begin{eqnarray}
(-1)^n\mathcal{L}_n&=&
\sum_{k=0}^{n}
\frac{1}{(n-k)!}\;\bar{\psi}^{\mu_1\ldots\mu_{n-k}}
[(-1)^{k}i\gamma^\sigma\nabla_\sigma-m_0]\psi_{\mu_1\ldots\mu_{n-k}}
\nonumber
\\
&&{}
-i\sum_{k=0}^{n-1}\frac{(-1)^{k}}{(n-k-1)!}\;
\Bigl(
\bar{\psi}^{\mu_1\ldots\mu_{n-k}}\nabla_{\mu_1}\psi_{\mu_2\ldots\mu_{n-k}}
+
\bar{\psi}^{\mu_2\ldots\mu_{n-k}}\nabla^{\mu_1}\psi_{\mu_1\ldots\mu_{n-k}}
\Bigr)
,
\label{kirdyk'}
\end{eqnarray}
where we have denoted
\begin{eqnarray}
\psi_{\mu_{k+1}\ldots\mu_n}=\frac{1}{k!}\gamma^{\mu_k}\ldots\gamma^{\mu_1}\psi_{\mu_1\ldots\mu_n}
\quad
\bar{\psi}_{\mu_{k+1}\ldots\mu_n}=
\frac{1}{k!}
\bar{\psi}_{\mu_1\ldots\mu_n}\gamma^{\mu_1}\ldots\gamma^{\mu_k}
,
\quad
\bar{\psi}_{\mu_1\ldots\mu_n}=
\psi_{\mu_1\ldots\mu_n}^+\gamma^0
.
&&
\end{eqnarray}
Thus we have constructed Lagrangian for antisymmetric massive
tensor-spinor field in terms of the basic field only.


\begin{thebibliography}{00}

\bibitem{reviews}
  M.~Vasiliev,
  ``Higher Spin gauge theories in various dimensions,"
  Fortsch.\ Phys.\ {\bf 52} (2004) 702
  [arXiv:hep-th/0401177];
  D.~Sorokin,
  ``Introduction to classical theory of higher spins,"
  AIP, Conf. Proc. {\bf 767} (2005) 172
  [arXiv:hep-th/0405069];
  N.~Bouatta, G.~Compare, A.~Sagnotti,
  ``An introduction to free higher-spin fields,"
  [arXiv:hep-th/0409068];
  X.~Bekaert, S.~Cnockert, C.~Iazeolla, M.~A.~Vasiliev,
  ``Nonliner higher spin theories in various dimensions,"
  [arXiv:hepth/0503128].

\bibitem{mixed1}
  L.~Brink, R.~R.~Metsaev,M.~A.~ Vasiliev,
  How massless are the massless fiemds in $AdS_d$,"
  Nucl.\ Phys.\ B {\bf 586} (2000) 183
  [arXiv:hep-th/0005136];
  K.~B.~Alkalaev, J.~V.~Shaynkman, M.~A.~Vasiliev,
  On the frame-like formulation of mixed symmetry massless fields in
  $(A)dS(d)$,"
  Nucl.\ Phys.\ B {\bf692} (2004) 363
  [arXiv:hep-th/0311164];
  Lagrangian formulation for free mixed-symmetry bosonic gauge
  fields in $(A)dS(d)$,"
  JHEP {\bf 0508} (2005) 069
  [arXiv:hep-th/0501108];
``Frame-like formulation for free mixed-symmetry bosonic massless
higher-spin fields in AdS(d),''
  arXiv:hep-th/0601225;
  R.~R.~Metsaev,
  ``Mixed symmetry massive fields in AdS(5),"
  Class.\ Quant.\ Grav.\ {\bf 22} (2005) 2777
  [arXiv:hep-th/0412311];
``Massless arbitrary spin fields in AdS(5),''
  Phys.\ Lett.\  B {\bf 531} (2002) 152-160
  [arXiv:hep-th/0201226];
``Arbitrary spin massless bosonic fields in d-dimensional anti-de Sitter
space,'' arXiv:hep-th/9810231;
``Massless mixed symmetry bosonic free fields in d-dimensional anti-de Sitter
space-time,''
  Phys.\ Lett.\  B {\bf 354} (1995) 78-84;
  K.~B.~Alkalaev,
  ``Mixed-symmetry gauge fields in $AdS_5$,"
  Theor.\ Math.\ Phys.\ {\bf 149} (2006) 1338
  [arXiv:hep-th/0501105].
E.D. Skvortsov,
``Mixed-Symmetry Massless Fields in Minkowski space Unfolded,''
  JHEP {\bf 0807} (2008) 004, arXiv:0801.2268 [hep-th]; ``Frame-like Actions for Massless
Mixed-Symmetry Fields in Minkowski space,''
  Nucl.\ Phys.\  B {\bf 808} (2009) 569, arXiv:0807.0903 [hep-th].










\bibitem{mixed2}
  X.~Bekaert, N.~Boulanger,
  ``Tensor gauge fields in arbitrary representations of $GL(D,R)$:
  duality and Poincare lemma,"
  Commun.\ Math.\ Phys.\ {\bf 245} (2004) 27
  [arXiv:hep-th/0208058];
  N.~Boulanger, C.~Iazeolla, P.~Sundell,
  ``Unfolding mixed-symmetry fields in AdS and the BMV conjecture:
  I. General formalism,"
  [arXiv:0812.3615];
  ``Unfolding mixed-symmetry fields in AdS and the BMV conjecture: II. Oscillator
  realization,"
  [arXiv:0812.4438].

\bibitem{mixed3}
  Yu.~M.~Zinoviev,
  ``On massive mixed symmetry tensor fields in Minkowski space and
  (A)dS,"
  [arXiv:hep-th/0211233];
  ``First order formalism for mixed symmetry tensor fields,"
  [arXiv:hep-th/0304067];
  ``First order formalism for massive mixed symmetry tensor fields
  in Minkowski and (A)dS spaces,"
  [arXiv:hep-th/0306292];
``On dual formulations of massive tensor fields,''
  JHEP {\bf 0510} (2005) 075
  [arXiv:hep-th/0504081];
  ``Toward frame-like gauge invariant formulation for massive mixed
  symmetry bosonic fields,"
  [arXiv:0809.3287].




\bibitem{mixed4}
F.~Bastinelli, F.~Benincasa, S.~Giombi,
``Worldline approach to vector and antisymmetric tensor fields,''
  JHEP {\bf 0504} (2005) 010 [arXiv:hep-th/0503155]; ``Worldline approach to
vector and antisymmetric tensor fields II,'' JHEP {\bf 0510} (2005) 114 [arXiv:hep-th/0510010];
K.~Hallowell and A.~Waldron,
``Constant curvature algebras and higher spin action generating  functions,''
  Nucl.\ Phys.\  B {\bf 724} (2005) 453
  [arXiv:hep-th/0505255];
F.~Bastianelli, O.~Corradini and E.~Latini,
  ``Higher spin fields from a worldline perspective,''
  JHEP {\bf 0702} (2007) 072
  [arXiv:hep-th/0701055];
``Spinning particles and higher spin fields on (A)dS backgrounds,''
  JHEP {\bf 0811} (2008) 054
 [arXiv:0810.0188 [hep-th]];
F.~Bastianelli and R.~Bonezzi,
``U(N) spinning particles and higher spin equations on complex manifolds,''
  arXiv:0901.2311 [hep-th];
F.~Bastianelli, O.~Corradini and A.~Waldron,
``Detours and Paths: BRST Complexes and Worldline Formalism,''
  arXiv:0902.0530 [hep-th].





\bibitem{08103467}
 I.~L.~Buchbinder, V.~A.~Krykhtin and L.~L.~Ryskina,
``BRST approach to Lagrangian formulation of bosonic totally antisymmeric
tensor fields in curved space,''
  arXiv:0810.3467 [hep-th].











\bibitem{massless-bos}
  A.~Pashnev and M.~Tsulaia,
``Description of the higher massless irreducible integer spins in the  BRST
approach,''
  Mod.\ Phys.\ Lett.\  A {\bf 13} (1998) 1853
  [arXiv:hep-th/9803207].
  X.~Bekaert, I.~L.~Buchbinder, A.~Pashnev and M.~Tsulaia,
``On higher spin theory: Strings, BRST, dimensional reductions,''
  Class.\ Quant.\ Grav.\  {\bf 21} (2004) S1457
  [arXiv:hep-th/0312252];
  A.~Fotopoulos, K.~L.~Panigrahi and M.~Tsulaia,
``Lagrangian Formulation Of Higher Spin Theories On AdS Space,''
  Phys.\ Rev.\  D {\bf 74} (2006) 085029
 [arXiv:hep-th/0607248].

\bibitem{B-BRST-Ads}
  I.~L.~Buchbinder, A.~Pashnev and M.~Tsulaia,
``Lagrangian formulation of the massless higher integer spin fields in  the
AdS background,''
  Phys.\ Lett.\  B {\bf 523} (2001) 338
  [arXiv:hep-th/0109067];
  I.~L.~Buchbinder, A.~Pashnev and M.~Tsulaia,
``Massless higher spin fields in the AdS background and BRST  constructions
for nonlinear algebras,''
  arXiv:hep-th/0206026.


\bibitem{boz-ferm}  I.~L.~Buchbinder, A.~V.~Galajinsky and V.~A.~Krykhtin,
``Quartet unconstrained formulation for massless higher spin
fields,''
  Nucl.\ Phys.\  B {\bf 779} (2007) 155-177
  [arXiv:hep-th/0702161];
  I.~L.~Buchbinder and A.~V.~Galajinsky,
  ``Quartet unconstrained formulation for massive higher spin fields,''
  JHEP {\bf 0811} (2008) 081
  [arXiv:0810.2852 [hep-th]].



\bibitem{0603212}
  I.~L.~Buchbinder, V.~A.~Krykhtin, L.~L.~Ryskina and H.~Takata,
  ``Gauge invariant Lagrangian construction for massive higher spin fermionic
  fields,''
  Phys.\ Lett.\  B {\bf 641} (2006) 386
  [arXiv:hep-th/0603212].



\bibitem{0703049}
  I.~L.~Buchbinder, V.~A.~Krykhtin and A.~A.~Reshetnyak,
  ``BRST approach to Lagrangian construction for fermionic higher spin   fields
  in AdS space,''
  Nucl.\ Phys.\  B {\bf 787} (2007) 211
  [arXiv:hep-th/0703049].




\bibitem{0608005}
  I.~L.~Buchbinder, V.~A.~Krykhtin and P.~M.~Lavrov,
  Nucl.\ Phys.\  B {\bf 762} (2007) 344-376
  [arXiv:hep-th/0608005];
I.~L.~Buchbinder and V.~A.~Krykhtin,
``Progress in Gauge Invariant Lagrangian Construction for Massive Higher Spin
Fields,''
  arXiv:0710.5715 [hep-th].



\bibitem{massive-bos}
  I.~L.~Buchbinder and V.~A.~Krykhtin,
``Gauge invariant Lagrangian construction for massive bosonic higher spin
fields in D dimensions,''
  Nucl.\ Phys.\  B {\bf 727} (2005) 537
  [arXiv:hep-th/0505092].
  I.~L.~Buchbinder and V.~A.~Krykhtin,
``BRST approach to higher spin field theories,''
  arXiv:hep-th/0511276.





\bibitem{0410215}
  I.~L.~Buchbinder, V.~A.~Krykhtin and A.~Pashnev,
  ``BRST approach to Lagrangian construction for fermionic massless higher
  spin fields,''
  Nucl.\ Phys.\  B {\bf 711} (2005) 367
  [arXiv:hep-th/0410215].





\bibitem{INT}
  I.~L.~Buchbinder, A.~Fotopoulos, A.~C.~Petkou and M.~Tsulaia,
``Constructing the cubic interaction vertex of higher spin gauge fields,''
  Phys.\ Rev.\  D {\bf 74} (2006) 105018
  [arXiv:hep-th/0609082];
  A.~Fotopoulos and M.~Tsulaia,
``Interacting Higher Spins and the High Energy Limit of the Bosonic String,''
  Phys.\ Rev.\  D {\bf 76} (2007) 025014
  [arXiv:0705.2939 [hep-th]];
  A.~Fotopoulos, N.~Irges, A.~C.~Petkou and M.~Tsulaia,
``Higher-Spin Gauge Fields Interacting with Scalars: The Lagrangian Cubic
Vertex,''
  JHEP {\bf 0710} (2007) 021
  [arXiv:0708.1399 [hep-th]];
  A.~Fotopoulos and M.~Tsulaia,
``Gauge Invariant Lagrangians for Free and Interacting Higher Spin Fields. A
Review of the BRST formulation,''
  arXiv:0805.1346 [hep-th].






\bibitem{0101201}
  C.~Burdik, A.~Pashnev and M.~Tsulaia,
  ``On the mixed symmetry irreducible representations of the Poincare group  in
  the BRST approach,''
  Mod.\ Phys.\ Lett.\  A {\bf 16} (2001) 731
  [arXiv:hep-th/0101201];
  P.~Y.~Moshin and A.~A.~Reshetnyak,
  ``BRST approach to Lagrangian formulation for mixed-symmetry fermionic
  higher-spin fields,''
  JHEP {\bf 0710} (2007) 040.
I.~L.~Buchbinder, V.~A.~Krykhtin and H.~Takata,
``Gauge invariant Lagrangian construction for massive bosonic mixed symmetry
higher spin fields,''
  Phys.\ Lett.\  B {\bf 656} (2007) 253
  [arXiv:0707.2181 [hep-th]].






\bibitem{9802097}
  R.~R.~Metsaev,
  ``Lowest eigenvalues of the energy operator for totally
  (anti)symmetric massless fields on the n-dimensional anti-de
  Sitter group,"
  Class.\ Quant.\ Grav.\ {\bf 11} (1994) L141;
  R.~R.~Metsaev,
  ``Free totally (anti)symmetric massless fermionic fields in
  d-dimensional anti-de Sitter space,"
  Class.\ Quant.\ Grav.\ {bf 14} (1997) L115
  [arXiv:hep-th/9707066];
  R.~R.~Metsaev,
  ``Fermionic fields in the d-dimensional anti-de Sitter spacetime,''
  Phys.\ Lett.\  B {\bf 419} (1998) 49
  [arXiv:hep-th/9802097].





\bibitem {BFV}E.S. Fradkin,  G.A. Vilkovisky, Quantization of relativistic
systems with constraints, Phys. Lett. B55 (1975) 224--226; I.A.
Batalin, G.A. Vilkovisky, Relativistic S-matrix of dynamical systems
with boson and fermion constraints, Phys. Lett. B69 (1977) 309--312;
I.A. Batalin, E.S. Fradkin, Operator quantization of relativistic
dynamical systems subject to first class constraints, Phys. Lett.
B128 (1983) 303.

\bibitem{bf}
I.A. Batalin, E.S. Fradkin, Operator quantization method and
abelization of dynamical systems subject to first class constraints,
Riv. Nuovo Cimento, 9, No~10 (1986) 1; I.A. Batalin, E.S. Fradkin,
Operator quantization of dynamical systems subject to constraints.
A further study of the construction, Ann. Inst. H. Poincare, A49
(1988) 145.






\bibitem{Francia}
  D.~Francia,
``Geometric Lagrangians for massive higher-spin fields,''
  Nucl.\ Phys.\  B {\bf 796} (2008) 77
  [arXiv:0710.5378 [hep-th]];
``Geometric massive higher spins and current exchanges,''
  Fortsch.\ Phys.\  {\bf 56} (2008) 800
  [arXiv:0804.2857 [hep-th]].

\end{thebibliography}
\end{document}